\documentclass[12pt]{article}
\usepackage[title]{appendix}
\usepackage{amsmath}
\usepackage{amsfonts}
\usepackage{amssymb}
\usepackage{graphicx}
\usepackage[round]{natbib}% \bibliography
\usepackage{float}
\usepackage{multirow}
\usepackage{tikz}
\usepackage{hyperref}
\usepackage{algorithm}
\usepackage[noend]{algpseudocode}
\usepackage{longtable}
\usepackage{makecell}
	
\usepackage{color, colortbl}
\usepackage{setspace} 
\usepackage{multicol}
\usepackage[margin=1in]{geometry}
\newcommand{\vecx}{\mathbf{x}}
\newcommand{\vecm}{\mathbf{m}}

\newcommand{\matT}{\textbf{T}}
\newcommand{\matS}{\textbf{S}}

\newcommand{\matD}{\textbf{D}}

\newcommand{\matX}{\mathbf{X}}

\newcommand{\vecM}{\mathbf{M}}
\newcommand{\matGamma}{\boldsymbol{\Gamma}}
\newcommand{\matSigma}{\boldsymbol{\Sigma}}
\newcommand{\vecmu}{\boldsymbol{\mu}}
\newcommand{\vecvartheta}{\boldsymbol{\vartheta}}

\newcommand{\matZero}{\boldsymbol{0}}

\newcommand{\vecdelta}{\boldsymbol{\delta}}

\newcommand{\vecX}{\mathbf{X}}

\definecolor{Gray}{gray}{0.9}

\usepackage{tabularx}
\usepackage{makecell}
\usepackage{array}
\usepackage{epsfig}
\usepackage{svg}
\usepackage{subfig}

\title{Hidden Markov Models for Multivariate \\ Panel Data}
\author{Mackenzie R.\ Neal, Alexa A.\ Sochaniwsky and Paul D.\ McNicholas}
\date{\small Department of Mathematics \& Statistics, McMaster University, ON, Canada.}

\begin{document}

\maketitle

\begin{abstract}
While advances continue to be made in model-based clustering, challenges persist in modeling various data types such as panel data. Multivariate panel data present difficulties for clustering algorithms because they are often plagued by missing data and dropouts, presenting issues for estimation algorithms. This research presents a family of hidden Markov models that compensate for the issues that arise in panel data. A modified expectation-maximization algorithm capable of handling missing not at random data and dropout is presented and used to perform model estimation. \newline

\end{abstract}

\section{Introduction}
Panel data arise from repeatedly collecting information on several subjects across several points in time. Research studies that utilize panel data are called longitudinal studies and are often found in clinical or sociological research. Difficulties may arise when analyzing panel data because missing data and dropouts are unavoidable and need to be properly accounted for.

One method of analysis for this type of data is model-based clustering, where the intent is to find groupings of observations in the data. A special case of model-based clustering uses dependent mixture models such as hidden Markov models (HMMs). HMMs are stochastic models that look to estimate an unobserved process based on an observed process, thus capturing underlying properties of the data. The unobserved process, or state sequence, is assumed to be a Markov chain, i.e., the next state is dependent on the past and present through the present state only. This dependent process makes these models useful in many applications including finance, epidemiology, sociology, and animal behaviour. 

Efforts have been made in regression models, specifically autoregressive and moving-average models \citep{hasan2009zero, sutradhar2003overview}, and in independent mixture models \citep{mcnicholas2010model}, to account for the unique longitudinal correlation structure. We present a modified expectation-maximization (EM) algorithm \citep{dempster1977maximum} utilizing a Cholesky-decomposed covariance matrix, similar to \citet{mcnicholas2010model} for Gaussian mixture models. The presented EM algorithm is further adapted to handle missing not at random (MNAR) data, similar to \citet{sportisse2021model} and \citet{speekenbrink2021ignorable}. \textcolor{black}{Thus, the proposed EM algorithm and family of HMMs is an improvement over traditional HMMs and their estimation procedures because: parsimonious covariance structures are possible; various types of non-random missingness are considered; and dropout is handled appropriately. These improvements are crucial to the modelling of panel data as all three issues --- parsimony, non-random missingness, and dropout --- are common in longitudinal studies.}

    The paper is structured in the following way. In Section 2, we discuss methods used in independent Gaussian mixture models to control for missing data, we then discuss HMMs and their estimation method. In Section~3, we introduce our proposed family of HMMs, present a method to handle informative missing data and dropouts, and provide a modified EM algorithm for estimation. In Section 4, we look at the performance of the new extensions on both simulated and real data. Lastly, in Section 5 we review the results and discuss recommendations for future work. Note that the notation throughout this paper is heavily influenced by \citet{maruotti2011mixed}, \citet{zucchini2009hidden}, and \citet{mcnicholas2010model}. \textcolor{black}{Note also that the methodology introduced herein is implemented in the  {\tt CDGHMM} package \citep{neal24a} for {\sf R} \citep{R23}.}

\section{Background}
In this section, we first discuss the methods used in Gaussian mixture models to improve model estimation in the presence of correlated observations and non-random missingness that we extend to HMMs. This is followed by a discussion of HMMs and their estimation.

\subsection{Covariance Structure for Panel Data}
%Longitudinal data cannot be treated as ``typical" data due to the correlation between measurements on each subject. To account for this in independent mixture models \citet{mcnicholas2010model} introduced the CDGMM family.

\subsubsection{\textcolor{black}{Modified Cholesky Decomposition}}\label{sec:modchol}
The Cholesky decomposition \citep{benoit1924note} is a matrix decomposition method whereby a real, positive-definite matrix can be decomposed into the product of a lower triangular matrix and its transpose. 
\cite{pourahmadi1999joint, pourahmadi2000maximum} introduces a modified Cholesky decomposition of covariance matrix $\matSigma$, given by
%\begin{equation*}
   $\matT \matSigma \matT' = \matD,$
%\end{equation*}
where $\matT$ is a unique unit lower triangular matrix and $\matD$ is a unique diagonal matrix with strictly positive diagonal entries. 
\cite{pourahmadi1999joint} shows that the elements of matrices $\mathbf{T}$ and  $\mathbf{D}$ can be considered to be generalized autoregressive parameters so that, based on $X_{t-1}, \ldots,X_1$, the least-squares predictor of $X_t$ can be written
\begin{equation*}
    \hat{X}_t = \mu_t + \sum_{s = 1}^{t-1} (- \phi_{ts})(X_s - \mu_s) + \sqrt{d_t \varepsilon_t},
\end{equation*}
where $\phi_{ts}$ is element $(t,s)$ of $\mathbf{T}$, $d_t$ is element $(t,t)$ of $\mathbf{D}$, i.e., the $t$th diagonal element of  $\mathbf{D}$, and $\varepsilon_t \sim \text{N}(0,1)$. 

\subsubsection{Gaussian Mixture Models for Panel Data}\label{sec:gmmpanel}
For panel data, \citet{mcnicholas2010model} use a Gaussian (independent) mixture model with a modified Cholesky decomposed covariance structure. For a $p$-dimensional random variable $\vecX$, the density of the $j$th component of a multivariate Gaussian mixture model with the modified-Cholesky decomposition is given by 
\begin{equation*}
    \phi(\vecx | \vecmu_j, (\matT_j' \matD_j^{-1} \matT_j)^{-1}) = \frac{1}{\sqrt{(2\pi)^p |\matD_j|} }\exp \left\{-\frac{1}{2}(\vecx - \vecmu_{j})' \matT_j' \matD_j^{-1} \matT_j (\vecx - \vecmu_j)\right\},
\end{equation*}
where $\matT_j$ is the $p \times p$ unit lower triangular matrix and $\matD_j$ is the $p \times p$ diagonal matrix that make up the modified Cholesky decomposition of $\matSigma_j$.

\citet{mcnicholas2010model} build a family of eight Gaussian mixture models by constraining $\matT_j$ and/or $\matD_j$, and this family is known as the Cholesky-decomposed Gaussian mixture model (CDGMM) family \citep{mcnicholas07}. Constraining $\mathbf{T}_j = \mathbf{T}$ suggests that all states have the same autoregressive relationship between points. Constraining $\mathbf{D}_j = \mathbf{D}$ suggests that all states have the same variability at each time point. The isotropic constraint $\mathbf{D}_j = d_j \mathbf{I}_p$, for $d_j\in\mathbb{R}^+$, where $\mathbf{I}_p$ denotes the $p\times p$ identity matrix, suggests that the variability at each point in time is the same. Thus, the combination of both constraints on $\matD_j$, i.e., $\mathbf{D}_j = d \mathbf{I}_p$, suggests that the variability across all states and all time points is the same. In Section \ref{models}, we introduce a family of Cholesky-decomposed HMMs.

\subsection{Missing Data} \label{missingback}
Missing data methods for model-based clustering is a well studied problem, including work by \citet{ghahramani1995learning, hunt2003mixture, eirola2014mixture}. The data are first partitioned into the observed and unobserved parts denoted $(\matX^o,\matX^m)$. By assuming the joint distribution of the missing and observed parts to be Gaussian, we can obtain the conditional distribution of the missing part given the observed part as
\begin{equation*}(\matX^m | \matX^o=\vecx^o) \sim \mathcal{N} (\vecmu_m +\matSigma_{mo}\matSigma_{oo}^{-1}(\vecx^o - \vecmu_o),\matSigma_{m|o})\end{equation*}
\citep{anderson2003introduction}. Based on this assumption, the following conditional expectations can be obtained:
\begin{equation*}\begin{split}
&\mathbb{E}(\matX_{it} | \vecx_{it}^o, j) = \left(\begin{matrix} \vecx_{it}^o \\
\vecmu_j^m + \matSigma_j^{mo}(\matSigma_j^{oo})^{-1}(\vecx_{it}^o
-\vecmu_j^o)\end{matrix}\right),\\  
&\mathbb{E}[(\matX_{it}-\vecmu_j)(\matX_{it}-\vecmu_j)' | \vecx_{it}^o, j] = \mathbb{V}\text{ar}(\matX_{it} | \vecx_{it}^o, j) + 
 [\mathbb{E}(\matX_{it} | \vecx_{it}^o,j) - \vecmu_j][\mathbb{E}(\matX_{it} | \vecx_{it}^o, j) - \vecmu_j]',\\
&\mathbb{V}\text{ar}(\matX_{it} | \vecx_{it}^o, j) = \left(\begin{matrix} \matZero & \matZero \\ \matZero & \matSigma_j^{mm} - \matSigma_j^{mo}(\matSigma_j^{oo})^{-1}\matSigma_j^{om} \end{matrix}\right).
\end{split}\end{equation*}
These expectations are used in the EM algorithm to account for missingness under a missing at random (MAR) scheme, as has been done to account for MAR data in model-based clustering and, more recently, in HMMs \citep{pandolfi2023hidden}. In this paper, we are interested in MNAR data, and there are many methods to handle such data within the model-based clustering paradigm \citep[e.g.,][]{du2023clustering, kuha2018latent, sportisse2021model}. \citet{sportisse2021model} allow for MNAR data by modeling the conditional distribution of missingness via a generalized linear model. We extend this method to HMMs as detailed in Section 3.2. Common methods for handling missing data in HMMs rely on imputing the missing values via the Viterbi algorithm \citep{popov2016training}, ignoring data gaps \citep{zucchini2009hidden}, missing at random algorithms \citep{pandolfi2023hidden}, and missing not at random algorithms wherein only state dependency is assumed \citep{speekenbrink2021ignorable}. Our method builds upon these methods by adding conditional mean imputation with various missingness mechanisms that allow for dependence on time, variables, and state. 

\subsection{Hidden Markov Models} 
\label{HMM}

One may opt for HMMs over an independent mixture model, as independent mixture models do not account for serial dependence in the observations. One way of allowing for serial dependence is to assume that the \textcolor{black}{underlying state sequence} is a Markov chain. The resulting model is known as a HMM. 

A HMM comprises of two processes, an unobserved \textcolor{black}{state sequence} and an observed state-dependent process. The simplest HMM for panel data can be defined as 
\begin{equation*}\begin{split}
    & P(C_{it} | \mathbf{C}^{(i,t-1)}) = P(C_{it} | C_{i,t-1}) \qquad\text{ for } i =1,\ldots,n, t = 2, 3, \ldots, T,\\ 
    & P(\vecX_{it} | \mathbf{X}^{(i,t-1)}, \mathbf{C}^{(it)}) = P(\vecX_{it} | C_{it} ) \ \text{ for } i = 1,\ldots,n, t = 1,\ldots, T,
\end{split}\end{equation*} 
where $\mathbf{C}^{(it)}$ represents the history of the unobserved \textcolor{black}{state sequence} $\{C_{it} : i = 1,\ldots, n, t = 1,2,\ldots,T \}$ with state space $C = 1,\ldots, m$, and $\mathbf{X}^{(it)}$ represents the history of the state-dependent process $\{\vecX_{it} : i = 1,\ldots, n, t = 1,2,\ldots,T \}$. Both processes, $C_{it}$ and $\vecX_{it}$, have unknown parameters. The state-dependent process has parameters specific to the assumed distribution; herein, we assume normality of the states and so our state-dependent parameters are $\vecmu_j$ and $\matSigma_j$. The unobserved \textcolor{black}{state sequence}, $C_{it}$, has an initial distribution $\boldsymbol{\delta}$, where $\boldsymbol{\delta}$ is a vector of length $m$ with its elements denoted by 
%\begin{equation*}
    $\delta_{ij} = P(C_{i1} = j),$
  and a transition matrix $\mathbf{\Gamma}$, denoted by 
\begin{equation*}
    \gamma_{ijk}(t) = P(C_{it} = j  | C_{i,t-1} = k)
\end{equation*}
for $j,k \in C$.  \textcolor{black}{We assume homogeneity, i.e., we assume that $\gamma_{ijk}(t) = \gamma_{jk}$ and $\delta_{ij} = \delta_j$; however, this assumption is not necessary for the work proposed herein \citep[see, e.g.,][]{fruhwirth2006finite}.}

\subsubsection{\textcolor{black}{Estimation}}

The likelihood equation arising from a HMM can be defined as 
\begin{equation*}
    L(\vecvartheta) = \prod_{i=1}^{n} \sum_{c_{i1} \in C} \hdots \sum_{c_{iT} \in C} \delta_{c_{i1}} \prod_{t=2}^T \gamma_{c_{i,t-1}c_{it}} \prod_{t=1}^T f(\vecx_{it}|c_{it}).
\end{equation*}
The computation of $L(\vecvartheta)$ is necessary but computationally expensive because, for each subject~$i$, the likelihood calculation requires a sum of $m^T$ terms \citep{maruotti2011mixed}. Using a recursive algorithm to perform estimation of $L(\vecvartheta)$ can greatly reduce the number of computations and, thus, reduce the chance of numerical instability. We focus on the forward-backward algorithm  proposed by \citet{baum1970maximization}, \citet{baum1972inequality}, and \citet{welch2003hidden}, known as the Baum-Welch algorithm, where the forward variables represent the probability of a sequence of events for subject $i$ ending up in state $j$ at time $t$. For $i = 1,\ldots,n$, $t = 1,\ldots,T$ and $j \in C$, the forward variable can be defined as
\begin{equation*}
    \alpha_{it}(j) = f(\vecx_{i1},\ldots,\vecx_{it}, C_{it} = j).
\end{equation*}
We can calculate $\alpha_{it}(j)$ through the following recursive process: 
\begin{enumerate}
    \item Calculate $\alpha_{i1}(j) = \mathbf{\delta}_j f(\vecx_{i1}| C_{i1} = j)$.
    \item For $t = 1,\ldots,T-1$: compute $\alpha_{i,t+1}(k) = \sum_{j=1}^m \alpha_{it}(j) \gamma_{jk} f(\vecx_{i,t+1} | C_{i,t+1} = k).$
\end{enumerate}
The likelihood can then be calculated via %by the following equation, 
%\begin{equation*}
    $L(\vecvartheta) = \prod_{i=1}^n \sum_{j=1}^m \alpha_{iT}(j).$
%\end{equation*}
The backward variable is defined as the probability of the sequence $(\vecx_{i,t+1}, \ldots., \vecx_{iT})$ occurring given that, at time $t$, the $i$th subject is in state $j$. That is, 
\begin{equation*}
    \beta_{it}(j) = f(\vecx_{i,t+1}, \ldots., \vecx_{iT} | C_{it} = j),
\end{equation*}
and the backward process is:
\begin{enumerate}
    \item $\beta_{iT}(j) = 1$.
    \item For $t = T-1,\ldots,1$: compute $\beta_{it}(j) = \sum_{k=1}^m \gamma_{jk} f(\vecx_{i,t+1} | C_{i,t+1} = k) \beta_{i,t+1}(k).$
\end{enumerate}
Using the forward and backward probabilities to calculate the conditional expectations of $u_{itj}$ and $v_{itjk}$ we get
\begin{align*}
    \hat{u}_{itj} & = \frac{\alpha_{it}(j) \beta_{it}(j) }{\sum_{j=1}^m \alpha_{it}(j) \beta_{it}(j)}, \qquad
    \hat{v}_{itjk}  =  \frac{\alpha_{i,t-1}(j) \hat{\gamma}_{jk} f(\vecx_{it} | C_{it} = k) \beta_{it}(k)}{\sum_{j,k=1}^m \alpha_{i,t-1}(j) \hat{\gamma}_{jk} f(\vecx_{it} | C_{it} = k) \beta_{it}(k)}.
\end{align*}
Together, the forward and backward probabilities, $\alpha_{it}(j)$ and $\beta_{it}(j)$, are used for model estimation via the Baum-Welch algorithm. The Baum-Welch algorithm is an EM algorithm for HMMs with a Markov chain that is homogeneous but does not have to be stationary, i.e., it is not assumed that $\boldsymbol{\delta} \mathbf{\Gamma} = \boldsymbol{\delta}$. The complete-data log-likelihood is
$$l(\vecvartheta)  =  \sum_{i = 1}^n \left\{ \sum_{j=1}^m
    u_{i1j} \log \delta_j +  \sum_{t = 2}^T \sum_{j=1}^m \sum_{k=1}^m v_{itjk} \log \gamma_{jk} + \sum_{t=1}^T \sum_{j=1}^m u_{itj} \log f(\vecx_{it} |  C_{it} = j ) \right\},$$
where $\vecvartheta$ denotes the vector of model parameters, $u_{itj}=1$ if $\vecx_{it}$ is in state $j$ and $u_{itj}=0$ otherwise, and $v_{itjk}=1$ if $\vecx_{i,t-1}$ is in state $j$ and $\vecx_{it}$ is in state $k$ and $v_{itjk}=0$ otherwise.  

For the E-step, the conditional expectations of the indicator variables, denoted by $\hat{u}_{itj}$ and $\hat{v}_{itjk}$, respectively, are calculated as
\begin{align*}
    \hat{u}_{itj} & = \frac{\alpha_{it}(j) \beta_{it}(j) }{\sum_{j=1}^m \alpha_{it}(j) \beta_{it}(j)}, \qquad
    \hat{v}_{itjk}  =  \frac{\alpha_{i,t-1}(j) \hat{\gamma}_{jk} f(\vecx_{it} | C_{it} = k) \beta_{it}(k)}{\sum_{j,k=1}^m \alpha_{i,t-1}(j) \hat{\gamma}_{jk} f(\vecx_{it} | C_{it} = k) \beta_{it}(k)}.
\end{align*}

For the M-step, we maximize the expected complete-data log-likelihood with respect to the initial distribution $\boldsymbol{\delta}$, transition matrix $\mathbf{\Gamma}$, and the assumed state-dependent distribution parameters --- in our case, the Gaussian distribution. This yields the following updates:
\begin{align}
    \hat{\delta}_j & = \frac{1}{n}\sum_{i=1}^n \hat{u}_{i1j},\qquad
    \hat{\gamma}_{jk} = \frac{1}{\sum_{i=1}^n \sum_{t=2}^T \sum_{k=1}^m \hat{v}_{itjk}}\sum_{i=1}^n \sum_{t=2}^T \hat{v}_{itjk},\nonumber\\
    \hat{\boldsymbol{\mu}}_j &= \frac{1}{\sum_{i=1}^n \sum_{t=1}^T \hat{u}_{itj}} \sum_{i=1}^n \sum_{t=1}^T \hat{u}_{itj} \vecx_{it},\nonumber\\
    \hat{\boldsymbol{\Sigma}}_j &=  \frac{1}{\sum_{i=1}^n \sum_{t=1}^T \hat{u}_{itj}} \sum_{i = 1}^n \sum_{t=1}^T \hat{u}_{itj} \left(\vecx_{it} -  \hat{\boldsymbol{\mu}}_j \right) \left(\vecx_{it} -  \hat{\boldsymbol{\mu}}_j \right)'.\label{eqn:thisone}
\end{align}

\textcolor{black}{The EM algorithm is commonly used to estimate HMM parameters. The traditional covariance update \eqref{eqn:thisone} may involve too many free parameters and facilitating parsimony might be desirable. In such cases, the proposed family of models discussed in Section~\ref{models}, which we shall call the Cholesky-decomposed Gaussian HMMs (CDGHMM) family, would allow for improved modelling of the covariance matrices. In Section \ref{method}, we introduce the CDGHMM family and further adjust the Baum-Welch algorithm to account for non-random missing data --- a common feature arising from longitudinal studies that is often overlooked.}

\section{Methodology}  \label{method}

\subsection{Models}\label{models}

\textcolor{black}{Herein, we present the CDGHMM family, which is analogous to the CDGMM family found in \citet{mcnicholas07} and \citet{mcnicholas2010model}. The HMM presented in Section~\ref{HMM} is modified as follows:
\begin{equation*}\begin{split}
    & P(C_{it} | \mathbf{C}^{(i,t-1)}) = P(C_{it} | C_{i,t-1}) \text{ for } i =1,\ldots,n, t = 2, 3, \ldots, T,\\ 
    & P(\vecX_{it} | \mathbf{X}^{(i,t-1)}, \mathbf{C}^{(it)}) = P(\vecX_{it} | C_{it} ) = \phi(\vecx | \vecmu_j, (\matT_j' \matD_j^{-1} \matT_j)^{-1}) \text{ for } i = 1,\ldots,n, t = 1,\ldots, T.
\end{split}\end{equation*} 
The complete-data log-likelihood of the proposed HMM is given by 
$$l(\vecvartheta)  = \sum_{i = 1}^n \left\{ \sum_{j=1}^m
    u_{i1j} \log \delta_j +  \sum_{t = 2}^T \sum_{j=1}^m \sum_{k=1}^m v_{itjk} \log \gamma_{jk}  + \sum_{t=1}^T \sum_{j=1}^m u_{itj} \log  \phi(\vecx_{it} | \vecmu_j, (\matT_j' \matD_j^{-1} \matT_j)^{-1}) \right\},$$
where $\matSigma_j$ has now been replaced with its modified Cholesky-decomposed form. The matrices $\matT_j$ and $\matD_j$ are constrained as discussed in Section~\ref{sec:gmmpanel} giving rise to the CDGHMM family (Table~\ref{tab:hmm_fam}).}
\begin{table}[ht]
\captionsetup{labelfont={color=black},font={color=black}}
\centering
\caption{The covariance matrix constraints and number of free covariance parameters for each member of the CDGHMM family.} 
\label{tab:hmm_fam}
\begin{tabular*}{0.9\textwidth}{@{\extracolsep{\fill}}ccccr}
 \hline
 \text{\textbf{Model}} & $\mathbf{T}_j$  & $\mathbf{D}_j$ & $\mathbf{D}_j$ & \text{\textbf{Free Cov. Parameters}} \\
 \hline
\text{EEA} & \text{Equal} & \text{Equal} & \text{Anisotropic} & $p(p-1)/2 + p$ \\
 
\text{VVA} & \text{Variable} & \text{Variable} & \text{Anisotropic} & $m(p(p- 1)/2) + mp$\\
  
\text{VEA} & \text{Variable} & \text{Equal} & \text{Anisotropic} & $m(p(p-1)/2) + p$ \\

\text{EVA} & \text{Equal} & \text{Variable} & \text{Anisotropic} & $p(p-1)/2 + mp$\\
 
\text{VVI} & \text{Variable} & \text{Variable} & \text{Isotropic} & $m(p(p-1)/2) + m$ \\
  
\text{VEI} & \text{Variable} & \text{Equal} & \text{Isotropic} & $m(p(p-1)/2) + 1$ \\

\text{EVI} & \text{Equal} & \text{Variable} & \text{Isotropic} & $p(p-1)/2 + m$\\

\text{EEI} & \text{Equal} & \text{Equal} & \text{Isotropic} & $ p(p-1)/2 + 1$\\
 \hline
% \bottomrule
\end{tabular*}
\end{table}

Parameter estimates for the least constrained model, VVA, and the most constrained model, EEI, are included below. These results are similar to those found in \citet{mcnicholas2010model}.

\subsubsection{VVA Model}
\label{VVAmodel}
The VVA model is unconstrained, i.e., neither the lower triangular matrix nor the diagonal matrix are constrained. The expected value of the complete-data log-likelihood is 
\begin{align*}
    Q  =  \sum_{i = 1}^n   & \left\{ \sum_{j=1}^m
    u_{i1j}  \log \delta_j +  \sum_{t = 2}^T \sum_{j=1}^m \sum_{k=1}^m v_{itjk} \log \gamma_{jk} \right\} 
     - \sum_{j=1}^m \frac{n_j p}{2} \log (2\pi) \\
     &\qquad\qquad - \sum_{j=1}^m \frac{n_j}{2} \log |\mathbf{D}_j| - \sum_{j=1}^m \frac{n_j}{2} \text{tr} \left\{ \matS_j \matT_j' \matD_j^{-1} \matT_j \right\},
\end{align*}
where $n_j = \sum_{i=1}^n\sum_{t=2}^{T} \hat{u}_{itj}$ and 
\[\matS_j = \frac{1}{n_j} \sum_{i=1}^{n} \sum_{t=1}^T \hat{u}_{itj} (\vecx_{it} - \vecmu_j) (\vecx_{it} - \vecmu_j)'. \]
Maximizing $Q$ with respect to $\pi_j$ and $\vecmu_j$ yields 
\begin{equation*}
    \hat{\pi}_j = \frac{n_j}{N} \hspace{10mm} \text{and} \hspace{10mm} \hat{\vecmu}_j = \frac{\sum_{i=1}^n\sum_{t=1}^{T} \hat{u}_{itj} \vecx_{it}}{\sum_{i=1}^n\sum_{t=1}^{T} \hat{u}_{itj}},
\end{equation*}
where $N = nT$. 
Noting that %Using some linear algebra results from Appendix \ref{appendix:linalg},  
%\begin{equation*}
    $\text{tr}\{\mathbf{S}_j \mathbf{T}_j' \mathbf{D}_j^{-1} \mathbf{T}_j\} = \text{tr}\{\mathbf{T}_j \mathbf{S}_j\mathbf{T}_j'\mathbf{D}_j^{-1} \},$
%\end{equation*}
the expected value of the complete-data log-likelihood can be written
\begin{equation}\label{eqn:q}
    Q = C - \frac{np}{2}\log 2\pi - \sum_{j = 1}^{m} \frac{n_j}{2} \log |\matD_j| - \sum_{j = 1}^{m} \frac{n_j}{2} \text{tr}\{\mathbf{T}_j \mathbf{S}_j \mathbf{T}_j'\mathbf{D}_j^{-1} \},
\end{equation}
where $C$ is a constant with respect to $\mathbf{T}_j$ and $\mathbf{D}_j$. %Using some of the matrix calculation results from Appendix \ref{appendix:matrix}, when 
Differentiating~\eqref{eqn:q} with respect to $\mathbf{T}_j$ and $\mathbf{D}_j$, respectively, we get the following score functions: 
\begin{align*}
    &S_1 (\mathbf{T}_j, \mathbf{D}_j) = \frac{\partial Q}{\partial \mathbf{T}_j} = -n_j  \mathbf{D}_j^{-1} \mathbf{T}_j \mathbf{S}_j,\\
    &S_2 (\mathbf{T}_j, \mathbf{D}_j) = \frac{\partial Q}{\partial \mathbf{D}_j} = \frac{n_j}{2} \left( \mathbf{D}_j - \mathbf{T}_j \mathbf{S}_j \mathbf{T}_j' \right).
\end{align*}
Let $\phi_{ig}^{(j)}$ denote the elements of $\matT_j$ such that 
\begin{equation*}
    \matT_j = \begin{pmatrix}
        1 & 0 & 0 & 0 & \hdots & 0\\
        \phi_{21}^{(j)} & 1 & 0 & 0 & \hdots & 0 \\
        \phi_{31}^{(j)} & \phi_{32}^{(j)} & 1 & 0 & \hdots & 0 \\
        \vdots & \vdots & & \ddots & & \vdots \\
        \phi_{p-1,1}^{(j)} & \phi_{p-1,2}^{(j)} & \hdots & \phi_{p-1,p-2}^{(j)} & 1 & 0 \\
        \phi_{p1}^{(j)} & \phi_{p2}^{(j)} & \hdots & \phi_{p,p-2}^{(j)} & \phi_{p,p-1}^{(j)} & 1 \\
    \end{pmatrix}
\end{equation*}
and  $\Phi_j = \{ \phi_{ig}^{(j)}\}$ for $i > g$, with $i,g\in \{1,\ldots,p\}$. Also, let LT$\{\cdot\}$ denote the lower triangle of a matrix. Solving $S_1(\hat{\matT}_j, \matD_j) \equiv $ LT$\{S_1(\hat{\Phi}_j, \matD_j) \} = 0$ for $\hat{\Phi}_j$ presents us with $p-1$ systems of linear equations. For the $1 \times 1 $ case, 
\begin{equation*}
    \frac{s_{11}^{(j)} \hat{\phi}_{21}^{(j)} }{d_2^{(j)}} + \frac{s_{21}^{(j)}}{d_2^{(j)}} = 0,
\end{equation*}
and so 
%\begin{equation*}
    $\hat{\phi}_{21}^{(j)} = -{s_{21}^{(j)}}/{s_{11}^{(j)}}$,
%\end{equation*}
where $s_{ig}^{(j)}$ denotes element $(i,g)$ of the matrix $\mathbf{S}_j$ and $d_i^{(j)}$ denotes the $i$th diagonal element of $\mathbf{D}_j$. For the general $(r-1) \times (r-1)$ case of system of equations,
\begin{equation*}
    \begin{pmatrix}
    \hat{\phi}_{r1}^{(j)} \\
    \hat{\phi}_{r2}^{(j)} \\
    \vdots \\
    \hat{\phi}_{r,r-1}^{(j)}
    \end{pmatrix} 
    = - \begin{pmatrix}
    s_{11}^{(j)} & s_{21}^{(j)} & \hdots & s_{r-1,1}^{(j)} \\
    s_{12}^{(j)} & s_{22}^{(j)} & \hdots & s_{r-1,2}^{(j)} \\
    \vdots & \vdots & \ddots & \vdots \\
    s_{1,r-1}^{(j)} & s_{2,r-2}^{(j)} & \hdots & s_{r-1,r-1}^{(j)}
    \end{pmatrix}^{-1} 
     \begin{pmatrix}
     s_{r1}^{(j)} \\
     s_{r2}^{(j)} \\
     \vdots \\
      s_{r,r-1}^{(j)} \\
     \end{pmatrix}
\end{equation*}
for $r = 2,\ldots,p$. Then, solving diag$\{S_2 (\hat{\mathbf{T}}_j, \hat{\mathbf{D}}_j)\} = 0$ gives 
%\begin{equation*}
    $\hat{\matD}_j = \text{diag}\{\hat{\matT}_j \matS_j \hat{\matT}_j'\}$.
%\end{equation*}
The resulting covariance update from the VVA model corresponds to the traditional covariance update \eqref{eqn:thisone}. 

\subsubsection{EEI Model}
\label{EEImodel}
For the EEI model, $\matT_j = \matT$ and $\matD_j = \matD = d \mathbf{I}_p$. Through a similar process as seen in the VVA derivation, the expected value of the complete-data log-likelihood for the EEI model can be written as 
\begin{equation*}
    Q = C + \sum_{j=1}^m + \frac{Np}{2} \log d^{-1} - \sum_{j=1}^m \frac{n_j d^{-1}}{2} \text{tr} \left\{ \matT \matS_j \matT' \right\},
\end{equation*}
where  $C$ is a constant with respect to $\matT$ and $d$, $n_j = \sum_{i=1}^n\sum_{t=2}^{T} \hat{u}_{itj}$, and 
\[\matS_j = \frac{1}{n_j} \sum_{i=1}^{n} \sum_{t=1}^T \hat{u}_{itj} (\vecx_{it} - \vecmu_j) (\vecx_{it} - \vecmu_j)'. \]
When differentiating the expected complete-data log-likelihood with respect to $\matT$ and $d^{-1}$ we get the following score functions:
\begin{align*}
    &S_1 (\matT, d) = \frac{\partial Q}{\partial \matT} = - \sum_{j=1}^m n_j d^{-1}\matT \matS_j,\\
    &S_2 (\matT, d) = \frac{\partial Q}{\partial d^{-1}} = \frac{N p d}{2} - \sum_{j=1}^m \frac{n_j}{2} \text{tr} \left\{ \matT \matS_j \matT' \right\}.
\end{align*}

Again, the lower triangular elements of $\matT$ are denoted by $\phi_{ig}$, and $\Phi = \left\{\phi_{ig} \right\}$ for $i>g$, with $i,g \in \{ 1,2,\ldots, p\}$. Solving LT$\left\{S_1 (\hat{\matT}, d) \right\}  \equiv [S_1 (\hat{\Phi}, d) ] = 0$, where $\hat{\Phi}$ consists of $p-1$ system of linear equations. In the case of a  $1 \times 1$ system, the solution is 
\begin{equation*}
    \sum_{j=1}^m n_j \left[ \frac{s_{11}^{(j)} \hat{\phi}_{21}}{d} + \frac{s_{21}^{(j)}}{d}  \right] = 0, 
\end{equation*}
and then solving for $ \hat{\phi}_{21}$ gives 
\begin{equation*}
     \hat{\phi}_{21} = - \frac{\sum_{j=1}^m \pi_j \left[s_{21}^{(j)} / d \right]}{\sum_{j=1}^m \pi_j \left[s_{11}^{(j)} / d \right]} = -\frac{\kappa^{21}}{\kappa^{11}},
\end{equation*}
where $\kappa^{ig} = \sum_{j=1}^m \pi_j \left[s_{ig}^{(j)} / d \right]$ for convenience. Following this pattern, the general solution for an $r-1$ system of equations is 

\begin{equation*}
    \begin{pmatrix}
    \hat{\phi}_{r1} \\
    \hat{\phi}_{r2} \\
    \vdots \\
    \hat{\phi}_{r,r-1}
    \end{pmatrix} 
    = - \begin{pmatrix}
    \kappa^{11} & \kappa^{21} & \hdots & \kappa^{r-1, 1} \\
    \kappa^{12} & \kappa^{22} & \hdots & \kappa^{r-1, 2} \\
    \vdots & \vdots & \ddots & \vdots \\
   \kappa^{1, r-1} & \kappa^{2, r-1} & \hdots & \kappa^{r-1, r-1} 
    \end{pmatrix}^{-1} 
     \begin{pmatrix}
    \kappa^{r1} \\
    \kappa^{r2} \\
     \vdots \\
     \kappa^{r,r-1}\\
     \end{pmatrix}
\end{equation*}
for $r = 2,\ldots,p$. Note that the $(r-1) \times (r-1)$ matrix above is symmetric because $\kappa_m^{ig} = \kappa_m^{gi}$. Then, solving $S_2(\hat{\matT}, \hat{d}) = 0$ for $\hat{d}$ gives
\begin{equation*}
    \hat{d} = \frac{1}{p}\sum_{j=1}^m\hat{\pi}_j\text{tr} \left\{ \hat{\matT} \matS_j \hat{\matT}' \right\}.
\end{equation*}

    Parameter estimation for the EEA and VEI models is detailed in Appendices~\ref{appEEA} and~\ref{appVEI}. 
   Parameter estimates for the VEA, EVI, EVA, and VVI are not included herein; however, the estimates under a a model-based clustering paradigm can be found in \citet{mcnicholas07} and \citet{mcnicholas2010model}.

\subsection{Missing Data} \label{missing}
As detailed in Section \ref{missingback}, conditional mean vectors and covariance matrices are obtained and used to perform imputation at each iteration of the EM algorithm, similar to \citet{pandolfi2023hidden} and \citet{eirola2014mixture}. To account for non-random missingness we introduce a missingness indicator variable as detailed below
\begin{equation*}
  M_{itj}=\begin{cases}
    1, & \text{if variable $j$ of $\vecX_{it}$ is unobserved},\\
    0, & \text{otherwise},
  \end{cases}
\end{equation*}
and assume conditional independence of $\vecX_{it}$ and $\vecM_{it}$ given the state, $f(\vecX_{it}, \vecM_{it} | C_{it}) = f(\vecX_{it}|C_{it}) P(\vecM_{it}|C_{it})$. Following \citet{sportisse2021model}, we calculate $f(\vecM_{it}|C_{it})$ via a generalized linear model with a probit link function and assume $f(\vecX_{it}|C_{it})$ to be Gaussian. In the presence of missing data, $f(\vecX_{it})$ is replaced with $f(\vecX_{it}^o)$, where $\vecX_{it}^{\{1 \times p\}}$ is the complete-data for individual $i$ at time $t$ and $\vecX_{it}^o$ is the observed data for individual $i$ at time $t$. If the data are completely unobserved for individual $i$ at time $t$, then $f(\vecX_{it}|C_{it}) =f(\vecX_{it}^o|C_{it})= 1$.

The parameterization of $f(\vecM_{it}|C_{it})$ varies based on the type of missingness. We allow for four types of non-random missingness, as detailed below in Sections~\ref{state_dep},\ref{state_var_dep}, \ref{state_time_dep}, and~\ref{state_var_time_dep}.

\subsubsection{State-Dependent Missingness} \label{state_dep}
When state-dependent missingness is assumed, we fit an intercept-only model to calculate $f(\vecM_{it}|C_{it}=c)$,
$$f(\vecM_{it} | C_{it}=c) = \prod_{j=1}^p \Phi(\alpha_c)^{m_{itj}}[1-\Phi(\alpha_c)]^{1-m_{itj}},$$
where $\alpha_c$ is the missingness parameter corresponding to state $c$.
\subsubsection{State- and Variable-Dependent Missingness}\label{state_var_dep}
When state- and variable-dependent missingness is assumed, we fit the following intercept-only model:
$$f(\vecM_{it} | C_{it}=c) = \prod_{j=1}^p \Phi(\alpha_{cj})^{m_{itj}}[1-\Phi(\alpha_{cj})]^{1-m_{itj}},$$
where $\alpha_{cj}$ is the missingness parameter corresponding to state $c$ and variable $j$.

\subsubsection{State- and Time-Dependent Missingness}\label{state_time_dep}
When we assume the effect of time on missingness is equal across states, we fit the following model:
$$f(\vecM_{it} | C_{it}=c) = \prod_{j=1}^p \Phi(\alpha_{c} + \beta_tt)^{m_{itj}}[1-\Phi(\alpha_{c} + \beta_tt)]^{1-m_{itj}},$$
where $\alpha_c$ is the missingness parameter corresponding to state $c$  and $\beta_t$ is the missingness parameter corresponding to time. With this parameterization of missingness, time can either be discrete or continuous. When we assume the effect of time on missingness varies by state, we fit the following model:
$$f(\vecM_{it} | C_{it}=c) = \prod_{j=1}^p \Phi(\alpha_{ct})^{m_{itj}}[1-\Phi(\alpha_{ct})]^{1-m_{itj}},$$
where $\alpha_{ct}$ is the missingness parameter for state $c$ at time $t$.

\subsubsection{State-, Variable-, and Time-Dependent Missingness}\label{state_var_time_dep}
When we assume the effect of time on missingness is equal across states but the effect of each variable differs across states, we fit the following model:
$$f(\vecM_{it} | C_{it}=c) = \prod_{j=1}^p \Phi(\alpha_{cj} + \beta_tt)^{m_{itj}}[1-\Phi(\alpha_{cj} + \beta_tt)]^{1-m_{itj}},$$
where $\alpha_{cj}$ is the effect of missingness of state $c$ and variable $j$, and $\beta_t$ is the missingess parameter for time. When we assume both the variable effect and time effect on missingness vary by state, we fit the following model
$$f(\vecM_{it} | C_{it}=c) = \prod_{j=1}^p \Phi(\alpha_{cjt})^{m_{itj}}[1-\Phi(\alpha_{cjt})]^{1-m_{itj}},$$
where $\alpha_{ctj}$ is the missingness parameter of state $c$ at time $t$ for variable $j$.
 
Once all the missingnesss parameters, $\alpha_1, \ldots, \alpha_m, \beta_1, \ldots, \beta_T$, are estimated, the observed likelihood is updated
\begin{equation*}
    L(\vecvartheta) = \prod_{i=1}^{n} \sum_{c_{i1} \in C} \hdots \sum_{c_{iT} \in C} \delta_{c_{i1}} \prod_{t=2}^T \gamma_{c_{i,t-1}c_{it}} \prod_{t=1}^T f(\vecx_{it}^o|c_{it})f(\vecm_{it}|c_{it}).
\end{equation*}
Again, we use the forwards and backwards probabilities to calculate the conditional expectations of $u_{itj}$ and $v_{itjk}$; therefore, both probabilities are updated to include $f(\vecM_{it} | C_{it})$.

The forward variable becomes
%\begin{equation*}
    $\alpha_{it}(j) = f(\vecx_{i1},\ldots,\vecx_{it},\vecm_{i1},\ldots,\vecm_{it},C_{it} = j),$
%\end{equation*}
and we calculate $\alpha_{it}(j)$ through the following recursive process: 

\begin{enumerate}
    \item Calculate $\alpha_{i1}(j) = \mathbf{\delta}_j f(\vecx_{i1}| C_{i1} = j)f(\vecm_{i1}| C_{i1} = j)$.
    \item For $t = 1,\ldots,T-1$: \newline
    compute $\alpha_{i,t+1}(k) = \sum_{j=1}^m \alpha_{it}(j) \gamma_{jk} f(\vecx_{i,t+1} | C_{i,t+1} = k)f(\vecm_{i,t+1} | C_{i,t+1} = k).$
\end{enumerate}

The backward variable becomes
%\begin{equation*}
    $\beta_{it}(j) = f(\vecx_{i,t+1}, \ldots., \vecx_{iT},\vecm_{i,t+1}, \ldots., \vecm_{iT} | C_{it} = j),$
%\end{equation*}
and the backwards process becomes:
\begin{enumerate}
    \item $\beta_{iT}(j) = 1$.
    \item For $t = T-1,\ldots,1$: \newline
    compute $\beta_{it}(j) = \sum_{k=1}^m \gamma_{jk} f(\vecx_{i,t+1} | C_{i,t+1} = k)f(\vecm_{i,t+1} | C_{i,t+1} = k) \beta_{i,t+1}(k).$
\end{enumerate}
%\begin{align*}
These updates, along with the updates discussed in Section \ref{missingback}, result in a Baum-Welch algorithm that can handle non-random missingness. In addition to missing data, panel data is often plagued by dropout, and as such further adjustments to the EM must be made.

\subsubsection{Dropout}
Following \citet{pandolfi2023hidden}, an indicator variable for dropout, $D_{it}$, and an absorbing state, $m+1$, are introduced, as follows:
\begin{align*}
f(D_{it} = d | C_{it} = j) =\begin{cases} 
1& \text{if } d=0 \text{ and } j=1,\ldots,m,\\ 
1&\text{if } d=1 \text{ and } j=m+1,\\
0& \text{otherwise},\end{cases}
\end{align*}%
where $D_{it}=1$ if an individual has dropped out and $D_{it}=0$ otherwise. Individuals can move into this state at any time $t>1$; however, once this state is reached further movement is prohibited. As such, in the presence of dropout, $\vecdelta$ and $\matGamma$ would take on the following forms:
\begin{equation*}
    \vecdelta=(\delta_1,\ldots,\delta_m,0), \qquad \matGamma = \left(\begin{matrix} \gamma_{11} & \ldots&\gamma_{1m} & \gamma_{1m+1} \\ 
\vdots & \ddots & \vdots & \vdots \\
\gamma_{m1}&\ldots &\gamma_{mm}& \gamma_{mm+1} \\
0 & \ldots& 0 & 1
 \end{matrix}\right).
\end{equation*}
 
Details of an EM for missing data and comments on an extension of this algorithm for dropout are found in Section \ref{EMcode}.

\subsubsection{Numerical stability}
 Numerical instability is a well known problem in parameter estimation of HMMs, as the likelihood is quick to go to zero or infinity \citep{zucchini2009hidden}. Oftentimes, a logarithmic transformation is used to overcome this issue; however, this approach cannot be used in the presence of dropout. Instead, we scale our forwards and backwards probabilities, i.e., 
\begin{equation*}
    \alpha_{it}(j) = \frac{f(\vecx_{i1},\ldots,\vecx_{it},\vecm_{i1},\ldots,\vecm_{it},C_{it} = j)}{\sum_{k=1}^{m+1} f(\vecx_{i1},\ldots,\vecx_{it},\vecm_{i1},\ldots,\vecm_{it},C_{it} = k)}
\end{equation*}
and
\begin{equation*}
    \beta_{it}(j) = \frac{f(\vecx_{i,t+1}, \ldots., \vecx_{iT},\vecm_{i,t+1}, \ldots., \vecm_{iT} | C_{it} = j)}{\sum_{k=1}^{m+1}f(\vecx_{i,t+1}, \ldots., \vecx_{iT},\vecm_{i,t+1}, \ldots., \vecm_{iT} | C_{it} = k)}.
\end{equation*}

\subsection{Model Estimation} \label{modestimation}
We present a modified EM algorithm for estimation of the CDGHMM family with missing data below, and discuss algorithm modifications that occur in the presence of dropout.

\subsubsection{CDGHMM EM Algorithm with Missing Data} \label{EMcode}
Algorithm~\ref{EMalgmiss} details the proposed model estimation algorithm in the presence of missing data.%
\begin{algorithm}
\caption{CDGHMM EM Algorithm with missing data.}\label{EMalgmiss}
\begin{algorithmic}[1]
\State initialize parameter estimates.
\State initialize $\hat{u}_{itj}$ and $\hat{v}_{itjk}$
\While{convergence criterion is not met}
\State update $$E(\matX_{it} | \vecx_{it}^o, j) = \left(\begin{matrix} \vecx_{it}^o\\ 
\vecmu_j^m + \matSigma_j^{mo}(\matSigma_j^{oo})^{-1}(\vecx_{it}^o
-\vecmu_j^o)\end{matrix}\right) $$
\State update $$\text{Var}(\matX_{it} | \vecx_{it}^o, j) = \left(\begin{matrix} \matZero & \matZero \\ \matZero & \matSigma_j^{mm} - \matSigma_j^{mo}(\matSigma_j^{oo})^{-1}\matSigma_j^{om} \end{matrix}\right)$$
\State update \begin{multline*}
E[(\matX_{it}-\vecmu_j)(\matX_{it}-\vecmu_j)' | \vecx_{it}^o, j] =\\ \text{Var}(\matX_{it} | \vecx_{it}^o, j) +
 [E(\matX_{it} | \vecx_{it}^o, j) - \vecmu_j][E(\matX_{it} | \vecx_{it}^o, j) - \vecmu_j]'
\end{multline*}
\State update $$\hat{u}_{itj} = \frac{\alpha_{it}(j)\beta_{it}(j)}{\sum_{j=1}^m \alpha_{it}(j) \beta_{it}(j)}$$
\State update $$\hat{v}_{itjk} = \frac{\alpha_{i,t-1}(j)\hat{\gamma}_{j,k} P_k(\matX_{it}^{o})  P_k(\vecM_{it}) \beta_{it}(k)}{\sum_{j,k=1}^m \alpha_{i,t-1}(j)\hat{\gamma}_{j,k} P_k(\matX_{it}^{o})  P_k(\vecM_{it}) \beta_{it}(k)} $$ 
\State update $$\hat{\gamma}_{jk} = \frac{1}{\sum_{i=1}^n \sum_{t=2}^T  \sum_{k=1}^K  \hat{v}_{itjk}}\sum_{i=1}^n \sum_{t=2}^T \hat{v}_{itjk}$$
\State update $$\hat{\delta}_j = \frac{1}{n}\sum_{i=1}^n \hat{u}_{i1j}$$
\State update $$\hat{\vecmu}_j = \frac{1}{\sum_{i=1}^n \sum_{t=1}^T \hat{u}_{ij}(t)} \sum_{i=1}^n \sum_{t=1}^T \hat{u}_{ij}(t) E(\matX_{it} | \vecx_{it}^o, j)$$
\State calculate $$\matS_j = \frac{1}{\sum_{i=1}^n \sum_{t=1}^T \hat{u}_{ij}(t)}\sum_{i=1}^n \sum_{t=1}^T   \hat{u}_{ij}(t)(E[(\matX_{it}-\vecmu_j)(\matX_{it}-\vecmu_j)' | \vecx_{it}^o, j])$$
\State update $\hat{\mathbf{T}}_j$, $\hat{\mathbf{D}}_j$ according to each model
\State update $ \hat{\boldsymbol{\Sigma}}_j^{-1} = \hat{\mathbf{T}}_j' \hat{\mathbf{D}}_j^{-1} \hat{\mathbf{T}}_j$
\State update $\hat{\boldsymbol{\alpha}}_j$ via a generalized linear model with a probit link on the appropriate $\zeta_{k}^{\text{MNAR}}$ matrix using $\hat{u}_{itj}$ as prior weights to the GLM.
\State check convergence criterion
\EndWhile
\State end while
\end{algorithmic}
\end{algorithm}

When we assume state-dependent missingness, the appropriate missing data matrix is
\begin{align*}
\zeta_{k}^{\text{MNAR}_z} &= \left(\begin{matrix} m_{111} & 1 \\ 
\vdots & \vdots \\
m_{nTj} & 1 
 \end{matrix}\right). 
\end{align*}
This matrix is modified slightly based on the type of missingness. 
When dropout is observed, the dimensions of $u_{itj}$, $v_{itjk}$, $\vecdelta$, and $\matGamma$ increase to accommodate the $m+1$ absorbing state. The mean vector and covariance matrix are unchanged by the dropouts. As such, steps 4--6 and 11--15 in Algorithm 1 occur for $j=1,\ldots, m$ and steps 7--10 occur for $j=1,\ldots, m+1$.

\subsubsection{Model Selection}

The integrated completed likelihood \citep{biernacki2000assessing} and the Bayesian information criterion  \citep[BIC;][]{schwarz1978estimating} are two of the model selection criterion options. The BIC can be written 
%\begin{equation*}
    $$\text{BIC} = 2l(\vecvartheta) - \rho \log N,$$
%\end{equation*}
where $l(\vecvartheta)$ is the maximized log-likelihood, $\rho$ is the number of free parameters, and $N = nT$. The ICL adds an entropy penalty to the BIC and can be approximated via 
\begin{equation*}
   \text{ICL} \approx \text{BIC} + 2 \sum_{i = 1}^n \sum_{t = 1}^T \sum_{j = 1}^m \text{MAP}\{\hat{u}_{itj}\} \log \hat{u}_{itj}, 
\end{equation*}
where  $\text{MAP}\{\hat{u}_{itj}\}$ is the maximum \textit{a~posteriori} classification, i.e.,
\begin{equation*}
    \text{MAP}\{\hat{u}_{itj}\} = \begin{cases} 
     1 & \text{if $j = \arg \matX_h  \{\hat{u}_{ith}\}$},\\
     0 & \text{otherwise.}
   \end{cases}
\end{equation*}
\textcolor{black}{Both the ICL and the BIC are considered as options for model selection; however, neither one is consistently better than the other. The BIC, ICL, and some other commonly-used criteria for selecting the number of states are investigated, and a summary of results on real and simulated data is given in Appendix~\ref{appCompSel}. Similar to \citet{pohle2017selecting} and \citet{celeux2008selecting}, we find the BIC has a tendency to select an unnecessarily large number of states \textcolor{black}{and the Akaike’s information criterion \citep[AIC;][]{akaike2011} exhibits similar behaviour.} We find ICL to have inconsistent behaviour, selecting the correct number of states on simulated data but an unnecessarily large number of states on real data. Of the methods tested, we find the average silhouette score \citep{Rousseeuw1987} performs the best in terms of selecting the correct number of states.}

\textcolor{black}{Regardless, we leave selection of the number of states as an area for future work, where one possible approach is the double-penalized likelihood \citep{hung2013, lin22, zou24} wherein penalties are added to the likelihood function and subsequent parameter estimators to avoid small groups and overfitting. Alternatively, an analysis of how much the chosen selection criterion improves while the number of states increases, in tandem with expert knowledge, could be used for selection of the number of states as discussed in \citet{pohle2017selecting}.}

\section{\textcolor{black}{Results}} \label{results}
\subsection{State Estimation and Performance Assessment}
The proposed EM algorithm is used to estimate $m$-state HMMs for a specified $m$ on both simulated and real data. \textcolor{black}{A local decoding algorithm is used to determine which state a given observation is most likely to be in at time $t$ via
\begin{equation}\label{eqn:prob}
P(C_{it}=j | \matX_i^{(T)} = \vecx_i^{(T)}) = \frac{P(C_{it}=j,\matX_i^{(T)} = \vecx_i^{(T)})}{P(\matX_i^{(T)} = \vecx_i^{(T)})}=\frac{\alpha_{it}(j)\beta_{it}(j)}{\sum_{j=1}^m \alpha_{iT}(j).},
\end{equation}
and the estimated state is selected as the state that maximizes the probability \eqref{eqn:prob} for a given observation.} Predicted states, \textcolor{black}{for some pre-specified $m$,} are compared to the known states for simulated data or a sensible categorical variable for the real datasets, and the misclassification rate can be used to assess performance. For comparison to the traditional HMM model, one can compare results from the seven constrained models to the VVA model. Additionally, comparison of the non-random missingness mechanisms can be made to the traditional MAR mechanism.

\subsection{Simulated Data}

\subsubsection{Simulation 1: \textcolor{black}{EEI Model}}\label{sim1}
The following simulation tests the performance of all members of the proposed CDGHMM family. We generate 250 datasets from an EEI model, where $T=5$ and $n\in\{100,500\}$, where $n$ represents the number of subjects and $T$ is the \textcolor{black}{sequence length} per subject. Three different transition matrices are used to \textcolor{black}{test how the degree of switching affects model performance:}
\begin{equation*}
    \matGamma_1=\left(\begin{matrix} 0.95 & 0.05 \\ 
0.05 & 0.95
 \end{matrix}\right),\qquad \matGamma_2 = \left(\begin{matrix} 0.5 & 0.5\\ 
0.5 & 0.5 
 \end{matrix}\right),\qquad \matGamma_3 = \left(\begin{matrix} 0.2 & 0.8\\ 
0.7 & 0.3
 \end{matrix}\right).
\end{equation*}
\noindent
Each dataset has the following parameters:
\begin{equation*}
\vecdelta=(0.5,0.5), \qquad
   \vecmu=\left(\begin{matrix} 3 & 4 & 5 & 10\\ 
        5&6&3&11
 \end{matrix}\right),\qquad 
   \matSigma=\left(\begin{matrix} 1 & 0& 0 & 0\\ 
        0 & 1& 0 & 0\\ 
        0& 0& 1 & 0\\ 
        0 & 0& 0 & 1
 \end{matrix}\right).
\end{equation*}
\setlength{\tabcolsep}{2pt} % Default value: 6pt
\begin{longtable}{ lccccccccc } 
\captionsetup{labelfont={color=black},font={color=black}}
\caption{Summary of simulation results for all models.}\label{tab:simres}\\
\hline
Model & $\matGamma$ & $N$ & \thead{Misclassification\\ rate: mean (s.d.)}& $\text{RMSE}(\hat{\matGamma})$ & $\text{RMSE}(\hat{\vecdelta})$ & $\text{RMSE}(\hat{\vecmu})$ & $\text{RMSE}(\hat{\matSigma})$& BIC & ICL\\
\hline
\multirow{6}{*}{} EEA & $\matGamma_1$ & 100 &0.01 (0.005) &0.02&0.05&0.06&0.07&$-6058.9$&$-16686.7$ \\ 
&  &  500  &0.01 (0.002) &0.01&0.03&0.03&0.03&$-29898.6$&$-82326.3$ \\ 
& $\matGamma_2$ &  100 & 0.14 (0.180)&0.17&0.17&0.45&0.41&$-6415.3$&$-12300.6$  \\ 
&  &  500  & 0.08 (0.140)&0.10&0.11&0.30&0.27&$-31591.4$&$-62467.6$ \\ 
& $\matGamma_3$ &  100 &0.4 (0.160) &0.53&0.31&0.87&0.74&$-6466.7$&$-10671.4 $\\ 
&  &  500& 0.39 (0.180)&0.50&0.29&0.84&0.72&$-31936.1$&$-50092.7 $\\ 
\hline
\multirow{6}{*}{} VVA & $\matGamma_1$  & 100 &0.01 (0.005) &0.02&0.05&0.06&0.07&$-6111.1$&$-16855.8$ \\ 
&  &500 & 0.01 (0.002) &0.01&0.03&0.03&0.03&$-29966.3$&$-82511.4$ \\ 
& $\matGamma_2$ &  100 & 0.13 (0.170)&0.18&0.15&0.40&0.40&$-6460.7$&$-13006.8 $\\ 
&  & 500 & 0.07 (0.110) &0.10&0.10&0.25&0.24&$-31638.2$&$-64970.1$  \\ 
& $\matGamma_3$ &  100 &0.41 (0.160) &0.57&0.28&0.83&0.79&$-6512.1$&$-13734.1 $\\ 
&  &  500 & 0.39 (0.180)&0.56&0.32&0.82&0.78&$-32014.5$&$ -63631.8$\\ 
\hline
\multirow{6}{*}{} VEA & $\matGamma_1$ & 100 & 0.01 (0.005)&0.02&0.05&0.06&0.07&$-6090.2$&$-16781.1 $\\ 
&  &  500& 0.01 (0.002) &0.01&0.03&0.03&0.03&$-29939.2$&$-82438.2$ \\
& $\matGamma_2$ &  100  & 0.12 (0.170)&0.17&0.15&0.40&0.38&$-6438.4$&$-12716.9$ \\
&  &  500  & 0.07 (0.110) &0.09&0.08&0.23&0.22&$-31607.2$&$-63249.6$ \\ 
& $\matGamma_3$ &  100 & 0.39 (0.170)&0.54&0.29&0.83&0.78&$-6492.5$&$-11228.1$ \\ 
&  &  500 & 0.35 (0.300)&0.51&0.28&0.037&0.77&$-31889.7$&$-53249.9 $\\ 
\hline
\multirow{6}{*}{} EVA & $\matGamma_1$ & 100& 0.01 (0.005)&0.02&0.05&0.06&0.07&$-6422.2$&$-20173.4 $\\ 
&  & 500 & 0.01 (0.002)&0.01&0.02&0.03&0.03&$-31610.3$&$-99583.0 $\\
& $\matGamma_2$ &  100 &0.04 (0.010) &0.04&0.05&0.07&0.08&$-6728.4$&$-17303.9$ \\ 
&  &  500 & 0.04 (0.004)&0.02&0.02&0.04&0.05&$-33154.7$&$-85088.7 $\\ 
& $\matGamma_3$ &  100 & 0.03 (0.008)&0.04&0.05&0.07&0.08&$-6643.1$&$-17874.4 $\\ 
&  &  500 & 0.03 (0.003)&0.02&0.02&0.04&0.04&$-32760.3$&$-88201.9$ \\ 

\hline
\multirow{6}{*}{} VVI & $\matGamma_1$ &  100 &0.01 (0.005) &0.02&0.05&0.06&0.07&$-6079.6$&$-16776.3$ \\ 
&  &  500 & 0.01 (0.002)&0.01&0.03&0.03&0.03&$-29925.7$&$-82424.5 $\\ 
& $\matGamma_2$ &  100 &0.04 (0.010) & 0.04&0.05&0.07&0.07&$-6404.7$&$-13283.3 $\\ 
&  &  500 &0.04 (0.004) &0.02&0.02&0.03&0.03&$-31552.4$&$-64752.5$ \\
& $\matGamma_3$ &  100 & 0.05 (0.080)&0.08&0.06&0.18&0.19&$-6325.5$&$-13743.7$ \\ 
&  &  500 & 0.03 (0.003)&0.02&0.02&0.03&0.03&$-31133.7$&$-68398.3$ \\ 
\hline
\multirow{6}{*}{}  VEI & $\matGamma_1$  &  100 &0.01 (0.005) &0.02&0.05&0.06&0.07&$-6074.5$&$-16749.9 $\\ 
  &  &  500 & 0.01 (0.002)&0.01&0.03&0.03&0.03&$-29918.9$&$-82403.0 $\\  
  & $\matGamma_2$ &  100 & 0.04 (0.010)&0.04&0.05&0.07&0.07&$-6399.4$&$-13243.9$ \\ 
&  &  500 &0.04 (0.004) &0.02&0.02&0.03&0.03&$-31545.6$&$-64700.5 $ \\ 
& $\matGamma_3$ &  100 & 0.08 (0.130)&0.12&0.08&0.28&0.29&$-6338.5$&$-13316.9 $\\ 
&  &  500 & 0.03 (0.003)&0.02&0.02&0.03&0.03&$-31126.9$&$-68365.9 $\\ 

\hline
\multirow{6}{*}{}  EVI & $\matGamma_1$  &  100 &0.01 (0.005) &0.02&0.05&0.06&0.07&$-6423.0$&$-20577.6 $\\ 
  &  & 500 &0.01 (0.002) &0.01&0.02&0.03&0.03&$-31723.2$&$-102052.8  $\\  
  & $\matGamma_2$ &  100 &0.04 (0.010) &0.04&0.05&0.07&0.07&$-6729.5$&$-17570.9 $\\ 
&  &  500 &0.04 (0.004) &0.02&0.02&0.04&0.04&$-33271.3$&$-87121.4$ \\ 
& $\matGamma_3$ &  100 &0.03 (0.008) &0.04&0.05&0.07&0.07&$-6644.8$&$-18219.9 $\\ 
&  &  500 &0.03 (0.003) &0.02&0.02&0.04&0.04&$-32873.9$&$-90218.9$ \\ 

\hline
\multirow{6}{*}{}  EEI & $\matGamma_1$   &  100&0.01 (0.005) &0.02&0.05&0.06&0.07&$-6043.19$&$-16650.8$\\ 
  &  &  500 &0.01 (0.002)&0.01&0.03&0.03&0.03&$-19878.3$&$-82285.4$  \\  
  & $\matGamma_2$  &  100&0.04 (0.01) &0.04&0.05&0.07&0.07&$-6368.1$&$-13106.4$ \\ 
&  &  500 & 0.04 (0.004)&0.02&0.02&0.03&0.03&$-31504.5$&$-64566.3 $\\
& $\matGamma_3$ &  100 &0.08 (0.13) &0.12&0.08&0.29&0.30&$-6311.2$&$-13091.0$ \\
&  & 500 &0.03 (0.003) &0.02&0.2&0.03&0.03&$-31086.3$&$-68234.8 $\\

\hline
\end{longtable}
%\end{table}

\textcolor{black}{From Table \ref{tab:simres}, under the specified simulation settings, all models perform well when there is infrequent switching. As the degree of switching increases, the misclassification rate for the EEA, VVA, and VEA models increases significantly.  As such, we show that the model with the traditional covariance update \eqref{eqn:thisone} --- which corresponds to the VVA model --- would fail in instances wherein the underlying covariance matrix is constrained and frequent switching occurs.}

\subsubsection{\textcolor{black}{Simulation 2: Panel Data}}\label{sim2}
\begin{figure}[ht]
\hspace{1.2cm}
\centering
\includegraphics[width=0.6\textwidth]{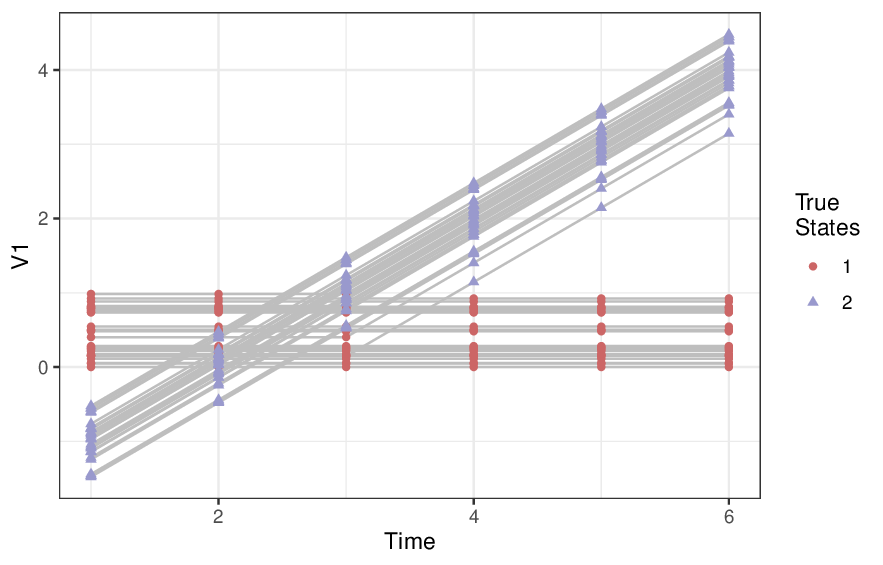}
\caption{True simulated states from Simulation 2.}
\label{fig:panel_ex}
\centering
\end{figure}

One motivation for the proposed work comes from clinical studies, wherein it may be of interest to have states represent trajectories, e.g., trajectories of an illness. In such instances, if an individual's trajectory is stable but begins to worsen at some point in the study, this decline in health would result in a switch to the state related to worsening health. We simulate this behaviour 250 times, where the true states can be fully described by variable~1 and variable~2 is random noise. The states are trajectories, and the true transition matrix is 
\begin{equation*}
 \matGamma = \left(\begin{matrix} 0.83 & 0.17\\ 
0 & 1
 \end{matrix}\right).
\end{equation*}
See Figure \ref{fig:panel_ex} for an example of the simulated data.

\begin{figure}[ht] 
    \centering
    \subfloat[\centering Most common VVI result.\label{fig:panelA}]{{\includegraphics[width=0.47\textwidth]{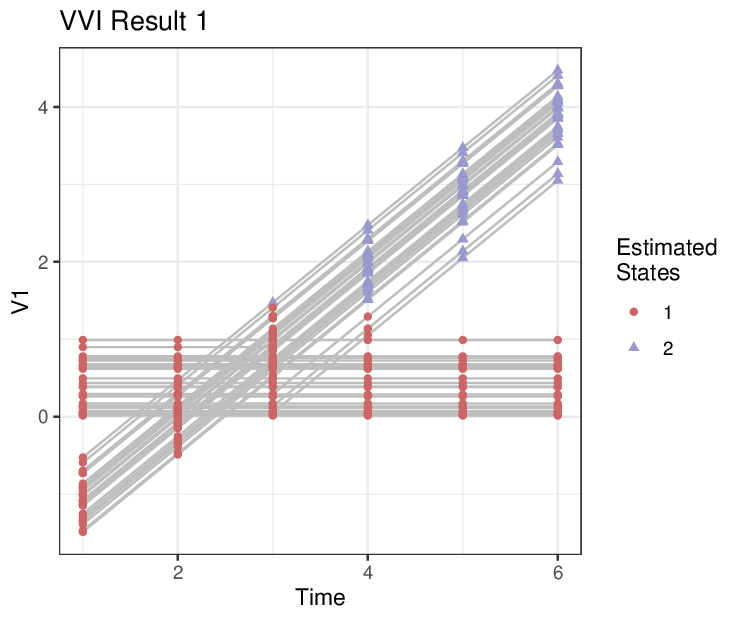}}}%
    \subfloat[\centering Second most common VVI result.\label{fig:panelB}]{{\includegraphics[width=0.47\textwidth]{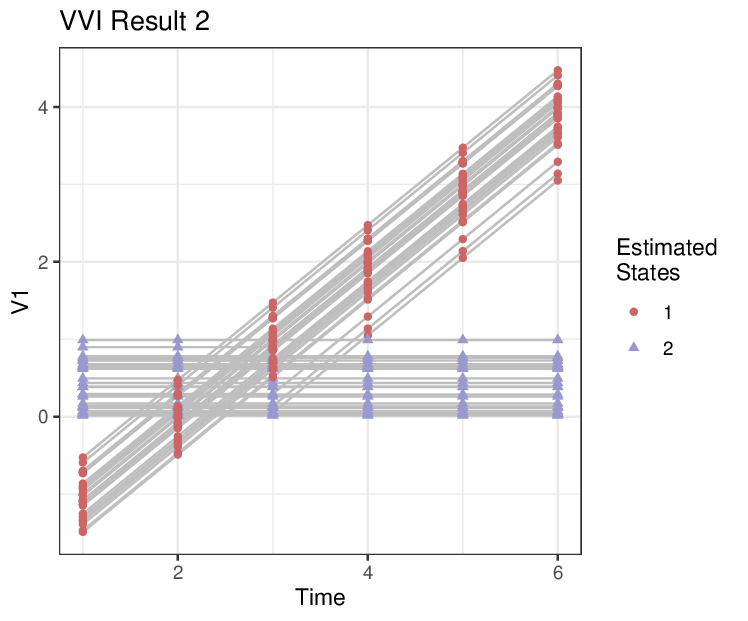}}}
    \newline
    \subfloat[\centering VVA result. \label{fig:panelC}]{{\includegraphics[width=0.47\textwidth]{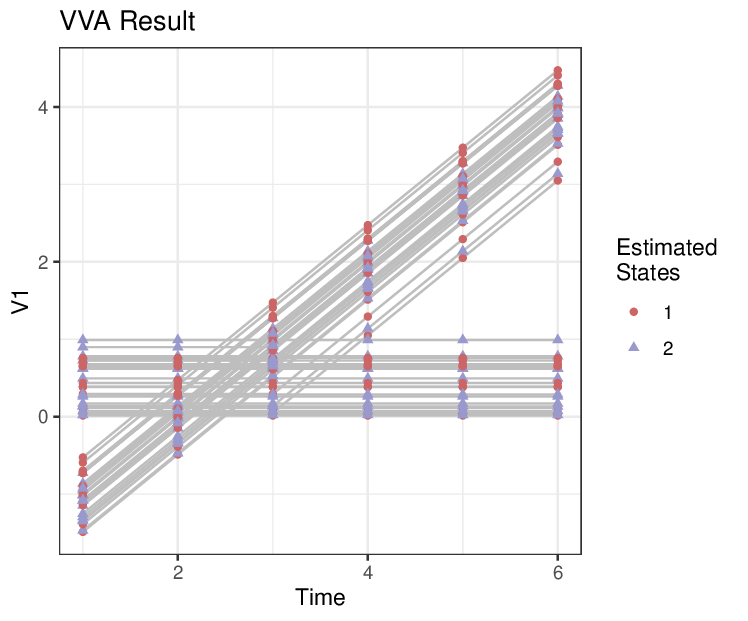}}} %
    \caption{Example of results from simulation 2 for the VVI and VVA models.}
      \label{fig:panel}
\end{figure}

State 1 corresponds to a stable state and state 2 corresponds to a state wherein the measurement for variable~1 continuously increases. Movement out of state 2 is not observed; however, five of the observational units move from state 1 to state 2 at time point 4. There are six time points and 60 observation units. We fit the proposed family to each simulated dataset and summarize the results in Table \ref{tab:panel}. Examples of the results are visualized in Figure \ref{fig:panel}. 
\begin{table}[H]
\centering
\caption{Summary of simulation 2 results.}
\label{tab:panel}
%\begin{tabular}{ lccccc } 
\begin{tabular*}{0.9\textwidth}{@{\extracolsep{\fill}}lccccc}
\hline
Model & \thead{Misclassification\\ rate: mean (s.d.)}& $\text{RMSE}(\hat{\matGamma})$ & $\text{RMSE}(\hat{\vecdelta})$ & BIC & ICL\\
\hline
EEA  &0.45 (0.04)& 0.12&0.07 &$-995.4$ & $-25359.3$\\ 
VVA  &0.46 (0.06)& 0.12 &0.10 &$-1003.3$ &$-30067.2$ \\ 
VEA &0.45 (0.04) &0.12 &0.07 &$-1000.3$ &$-25342.4$ \\ 
EVA &0.44 (0.06) &0.12 &0.10 & $-997.5$& $-30278.9$\\ 
VVI & 0.21 (0.07) &0.02 &0.4 &$-1363.7$ &$-32288.3$ \\ 
VEI &0.26 (0.07)&0.02 &0.5 & $-1360.5$& $-38472.6$\\ 
EVI &0.26 (0.07)&0.02 &0.5 & $-1388.5$& $-54406.4$\\ 
EEI &0.26 (0.07) & 0.02& 0.5& $-1354.7$ & $-38497.4$\\ 
\hline
\end{tabular*}
\end{table}
%\begin{figure}[ht] 
%    \centering
%    \subfloat[\centering Most common VVI result.\label{fig:panelA}]{{\includegraphics[width=0.47\textwidth]{vvi_trajsim.eps}}}%
%    \subfloat[\centering Second most common VVI results.\label{fig:panelB}]{{\includegraphics[width=0.47\textwidth]{vvi_trajsim_res2}}}
%    \caption{Summary of HMM results from simulation 2 for the VVI and VVA models.}
%    \label{fig:panel}
%    \end{figure}
%    \begin{figure}[H]
%    \ContinuedFloat\centering
%    \subfloat[\centering VVA results \label{fig:panelC}]{{\includegraphics[width=0.47\textwidth]{VVA_trajsim}}} %
%    \caption{Summary of results from simulation 2 for the VVI and VVA models.}
%      \label{fig:panel}
%\end{figure}
Every iteration of the simulation produces VVA results similar to those found in Figure~\ref{fig:panelC}. The VVI model results in minor variations of two solutions, and examples of each are included in Figures~\ref{fig:panelA} and~\ref{fig:panelB}. We see that not only does the VVA model have the worst predictive performance (Table~\ref{tab:panel}) but results in nonsense states (Figure~\ref{fig:panelC}). Whereas, the VVI model either produces the desired outcome or an alternative but sensible solution. Because the VVA model does not arrive at the desired outcome or any sensible outcome, this simulation suggests that the traditional covariance update \eqref{eqn:thisone} could fail when applied to data with unique correlation structures, and by constraining the covariance matrix we are better able to capture states that are defined by trajectories.

\subsubsection{Simulation 3: Missing Data Mechanisms} \label{sim3}
To measure performance of three missing mechanisms (MAR, state-dependent missingness, and state- and variable-dependent missingness), we simulate data 250 times under various parameter settings (Table \ref{tab:missing_param}) and fit a VVA model with each missingness mechanism to each simulated dataset. 
\begin{table}[!ht]
\centering
\caption{Summary of simulation 3 settings. }
\label{tab:missing_param}
\begin{tabular}{ lccccccc } 
\hline
$m$ &$\vecdelta$& $\vecmu$ & $\matGamma$ & \thead{Missing \\ parameters}\\
\hline
2 &$\vecdelta=(0.2,0.8,0)$ & $\left(\begin{matrix}
    5 & 4 & 5 &10\\
    5 & 6.5 &3 &10
\end{matrix}\right)$ &  $\left(\begin{matrix} 0.65 & 0.05 &0.3\\ 
0.25 & 0.7 & 0.05\\
0 &0 & 1 
 \end{matrix}\right)$  & \thead{$p_{\text{miss}}= (0.1,0.3,0.5)$ \\ $m_{\text{miss}} = (0.7,0.3)$ \\ $v_{\text{miss}} = (0,0.6,0.4,0)$}\\ 
\hline
3 &  $\vecdelta=(0.3,0.6,0.1,0)$&$\left(\begin{matrix}
    5 & 4 & 5 &10\\
    5 & 6.5 &3 &10 \\
     5 & 3 &2 &7
\end{matrix}\right)$ &  $\left(\begin{matrix} 0.45 & 0.15 &0.30& 0.10\\ 
0.2 & 0.7 & 0.05 & 0.05 \\
0.15 &0 & 0.7 & 0.15 \\
0 & 0 & 0 &1
 \end{matrix}\right)$ & \thead{$p_{\text{miss}}= (0.1,0.3,0.5)$ \\ $m_{\text{miss}} = (0.7,0,0.3)$ \\ $v_{\text{miss}} = (0,0.5,0.4,0.1)$} \\ 
\hline
\end{tabular}
\end{table}

Two sample sizes are tested for each simulation, i.e., $n\in\{100, 500\}$ and $T=5$, where $n$ represents the number of units and $T$ represents the \textcolor{black}{sequence length}. All states, in each simulation, have the covariance structure
$$\matSigma = \left(\begin{matrix} 1 & 0.5 &0& 0.25\\ 
0.50 & 1 & 0.5 & 0 \\
0 &0.5 & 1 & 0 \\
0.25 & 0 & 0 &1
 \end{matrix}\right).$$
\begin{table}[ht]
\centering
\caption{Summary of simulation 3 results for missing data mechanisms when $m=2$. }
\label{tab:missing1}
\begin{tabular*}{0.9\textwidth}{@{\extracolsep{\fill}}lccccccc}
\hline
\thead{Missing \\ mechanism} & $p_{\text{miss}}$ & $n$ & \thead{Misclassification\\ rate: mean (s.d.) }& $\text{RMSE}(\hat{\matGamma})$ & $\text{RMSE}(\hat{\vecdelta})$ & $\text{RMSE}(\hat{\vecmu})$ & $\text{RMSE}(\hat{\matSigma})$\\
\hline
\multirow{6}{*}{} MAR & 0.1 &  100 & 0.013 (0.006) & 0.027& 0.001& 0.075& 0.089\\ 
& 0.1 &  500 & 0.012 (0.003) &0.013 & 0.0002&0.035&0.039 \\ 
& 0.3 &  100 & 0.046 (0.012) & 0.031&0.002 &0.086&0.098 \\ 
& 0.3 &  500 & 0.044 (0.005) & 0.016& 0.0009&0.041&0.045 \\ 
& 0.5 &  100 & 0.120 (0.030) & 0.050&0.005 &0.129&0.143 \\ 
& 0.5 &  500 & 0.110 (0.012) &0.032 &0.004 &0.065&0.060  \\ 
\hline
\multirow{6}{*}{} State & 0.1 & 100 &  0.009 (0.005) & 0.026& 0.001&0.074&0.088 \\ 
& 0.1 &  500 & 0.009 (0.002) &0.012 &0.0002 &0.034&0.039 \\ 
& 0.3 &  100 & 0.024 (0.009) & 0.027&0.001 &0.082&0.097 \\ 
& 0.3 &  500 & 0.023 (0.004) & 0.012&0.0002 &0.037&0.420  \\ 
& 0.5 &  100 & 0.043 (0.013) & 0.030&0.001 &0.113&0.137 \\ 
& 0.5 &  500 & 0.039 (0.005) &0.014 &0.0004 &0.045&0.052 \\ 
\hline
\multirow{6}{*}{}State \&   & 0.1 &  100 & 0.009 (0.005) & 0.026 &0.001 &0.074& 0.088\\ 
Variable & 0.1 &  500 &  0.008 (0.002) & 0.012&0.0002 &0.034&0.039 \\ 
& 0.3 & 100 &  0.024 (0.008) & 0.027&0.001 &0.082&0.097 \\ 
& 0.3 &  500 & 0.023 (0.004) & 0.012&0.0002 &0.036&0.042 \\ 
& 0.5 & 100 &  0.039 (0.013) & 0.029&0.001 &0.129&0.129 \\ 
& 0.5 &  500 & 0.037 (0.005) &0.013 &0.0002 &0.041&0.051 \\ 
\hline
\end{tabular*}
\end{table}

From Tables \ref{tab:missing1} and \ref{tab:missing2}, we see that the proportion of missingness has a negative effect on predictive performance for all methods. However, in all cases, the performance of the MAR method declined more than the methods that account for non-random missingness. This is particularly true in Table \ref{tab:missing2}, when $p_{\text{miss}} = 0.5$: the misclassification rate is 0.29 for MAR but is only 0.09 for the state-dependent missingness mechanism. Large differences in performance between the state-dependent mechanism and the state- and variable-dependent mechanism are not observed, as is expected due to the global variable missingness parameters.
\begin{table}[ht]
\centering
\caption{Summary of simulation 3 results for missing data mechanisms when $m=3$. }
\label{tab:missing2}
\begin{tabular*}{0.9\textwidth}{@{\extracolsep{\fill}}lccccccc} 
\hline
\thead{Missingness\\ mechanism} & $p_{\text{miss}}$ & $n$ & \thead{Misclassification \\ rate: mean (s.d.) }& $\text{RMSE}(\hat{\matGamma})$ & $\text{RMSE}(\hat{\vecdelta})$ & $\text{RMSE}(\hat{\vecmu})$ & $\text{RMSE}(\hat{\matSigma})$\\
\hline
\multirow{6}{*}{} MAR & 0.1 &  100 & 0.08 (0.021) & 0.09&0.011 &1.48&0.12 \\ 
& 0.1 &  500 & 0.07 (0.007) &0.02 &0.001 &1.51& 0.05\\ 
& 0.3 &  100 & 0.16 (0.035) & 0.11 & 0.021& 1.45& 0.16 \\ 
& 0.3 &  500 & 0.14 (0.010) &0.06 & 0.007 & 1.50& 0.06\\ 
& 0.5 &  100 & 0.29 (0.074) & 0.15& 0.033&1.39&0.26 \\ 
& 0.5 &  500 & 0.25 (0.014) & 0.11& 0.030&1.46& 0.10 \\ 
\hline
\multirow{6}{*}{} State  &0.1 &100 &  0.05 (0.017) &0.08&0.008 &1.49& 0.11\\ 
& 0.1 &  500 & 0.05 (0.005) &0.02 &0.0004 &1.51& 0.05\\ 
& 0.3 &  100 & 0.05 (0.013) & 0.08 & 0.012 &1.47& 0.17 \\ 
& 0.3 &  500 & 0.05 (0.005) & 0.04& 0.001&1.50 &0.06 \\ 
& 0.5 &  100 & 0.09 (0.100) & 0.09& 0.015&1.45& 0.22\\ 
& 0.5 &  500 & 0.05 (0.005) & 0.02& 0.0004&1.49& 0.06\\ 
\hline
\multirow{6}{*}{}State \&  & 0.1 &100 & 0.06 (0.018) &0.08 & 0.009 & 1.49& 0.11\\ 
Variable & 0.1 &  500 & 0.05 (0.005) & 0.02& 0.00004& 1.51& 0.05\\ 
& 0.3 & 100 & 0.05 (0.014) & 0.08& 0.013&1.48& 0.13 \\ 
& 0.3 &  500 & 0.05 (0.005) &0.04 & 0.001&1.51 & 0.01\\ 
 & 0.5 & 100 &  0.10 (0.110) &0.09 &0.017 &1.46&0.23\\ 
& 0.5 &  500 & 0.05 (0.020) & 0.02&0.0009 &1.51&0.07 \\ 
\hline
\end{tabular*}
\end{table}

\subsubsection{\textcolor{black}{Simulation 4: CDGHMM family and Missing Data Mechanisms}} \label{sim4}
To test the performance of the CDGHMM family with the proposed missingness mechanisms, we simulate data 250 times under various parameter settings, detailed in Table \ref{tab:missing_param2}. For each simulated dataset, $n=500$, $T=5$, and $\matSigma_1=\matSigma_2=\matSigma_3 = \mathbf{I}_4$.  We fit the CDGHMM family under two missingness mechanisms, MAR and state and variable dependent missingness, to each simulated dataset and summarize the results in Table \ref{tab:s4m2} and Table \ref{tab:s4m3}. 

\begin{table}[ht]
\centering
\caption{Summary of settings for simulation 4. }
\label{tab:missing_param2}
\begin{tabular}{ lccccccc } 
\hline
$m$ &$\vecdelta$& $\vecmu$ & $\matGamma$ & \thead{Missing \\ parameters}\\
\hline
2 &$\vecdelta=(0.5,0.5)$ & $\left(\begin{matrix}
    3 & 5 & 3 &10\\
    5 & 4 &3 &11
\end{matrix}\right)$ &  $\left(\begin{matrix} 0.5 & 0.5 \\ 
0.5 & 0.5 
 \end{matrix}\right)$  & \thead{$p_{\text{miss}}=0.30$ \\ $m_{\text{miss}} = (0.8,0.2)$ \\ $v_{\text{miss, }m=1} = (0.8,0.2,0,0)$\\ $v_{\text{miss, }m=2} = (0,0,0.5,0.5)$}\\
\hline
3 &  $\vecdelta=(0.33,0.33,0.33)$&$\left(\begin{matrix}
    5 & 6 & 5 &10\\
    5 & 6 &3 &10.5 \\
     5 & 5.5 &2 &9
\end{matrix}\right)$ &  $\left(\begin{matrix} 0.2 & 0.4 &0.4\\ 
0.4 & 0.2 & 0.4 \\
0.4 &0.4 & 0.2 
 \end{matrix}\right)$ & \thead{$p_{\text{miss}}=0.30 $ \\ $m_{\text{miss}} = (0.8,0,0.2)$ \\ $v_{\text{miss, }m=1} = (0,0.8,0.2,0)$ \\  $v_{\text{miss, }m=2} = (0,0,0,0)$\\ $v_{\text{miss, }m=3} = (0,0,0.5,0.5)$}\\
\hline
\end{tabular}
\end{table}
From Table \ref{tab:s4m2}, we see that for all members of the CDGHMM family, controlling for non-random missingness significantly boosts performance over the MAR method. Additionally, we see that although some models (EEA, VVA, VEA, EVA) perform far worse under the MAR scheme than others (VVI, VEI, EVI, EEI), the non-random missingness mechanism is able to boost all models to approximately the same performance. 
\begin{table}[!h]
\centering
\caption{Summary of simulation 4 results when $m=2$. }
\label{tab:s4m2}
\begin{tabular*}{0.9\textwidth}{@{\extracolsep{\fill}}lcccccc} 
\hline
Model & \thead{Missingness \\ mechanism} & \thead{Misclassification\\ rate: mean (s.d.) }& $\text{RMSE}(\hat{\matGamma})$ & $\text{RMSE}(\hat{\vecdelta})$ & $\text{RMSE}(\hat{\vecmu})$ & $\text{RMSE}(\hat{\matSigma})$\\
\hline
\multirow{2}{*}{}EEA & MAR & 0.30 (0.12) &0.05 &0.17 & 0.34&0.26 \\ 
& \thead{State \& \\ Variable}& 0.08 (0.013) & 0.23& 0.06&0.08& 0.06\\ 

\hline
\multirow{2}{*}{}VVA& MAR & 0.30 (0.11) &0.25 & 0.21& 0.35& 0.30\\ 
& \thead{State \& \\ Variable}& 0.09 (0.016) & 0.05& 0.07&0.09& 0.10\\ 

\hline
\multirow{2}{*}{}VEA    & MAR & 0.31 (0.13) &0.24 & 0.18& 0.34& 0.30\\ 
& \thead{State \& \\ Variable}& 0.08 (0.014) &  0.05&0.06 &0.08&0.08 \\ 
 \hline
\multirow{2}{*}{}EVA    & MAR & 0.28 (0.12) & 0.23&0.20 & 0.32& 0.25\\  
& \thead{State \& \\ Variable}& 0.08 (0.014) & 0.05& 0.06&0.09&0.08 \\ 
 \hline
\multirow{2}{*}{}VVI    & MAR & 0.18 (0.014) & 0.11& 0.12&0.13 & 0.10\\  
& \thead{State \& \\ Variable}& 0.08 (0.04) & 0.05&0.06 &0.08& 0.08\\ 
 \hline
\multirow{2}{*}{}VEI   & MAR & 0.17 (0.030) & 0.10&0.10 & 0.12&0.09 \\ 
& \thead{State \& \\ Variable}& 0.08 (0.013) & 0.04&0.06 &0.08& 0.08\\ 
 \hline
\multirow{2}{*}{}EVI    & MAR & 0.16 (0.012) &0.09 & 0.10& 0.10& 0.05\\ 
 & \thead{State \& \\ Variable}& 0.08 (0.012) &0.05 & 0.06&0.08& 0.05\\ 
 \hline
\multirow{2}{*}{}EEI   & MAR & 0.17 (0.03) & 0.10& 0.10& 0.11& 0.07\\ 
& \thead{State \& \\ Variable}& 0.08 (0.013)  &0.05 &0.06 &0.08& 0.05\\ 
\hline
\end{tabular*}
\end{table}

In Table \ref{tab:s4m3}, we see that the state- and variable-dependent missingness mechanism is able to boost performance for all CDGHMM family members over the MAR method. However, in this case we do not see all models performing approximately the same under the non-random missingness mechanism. Rather, the EEI and EVI models under a state- and variable-dependent missingness scheme perform the best, with misclassifications rates 14 percentage points lower than the misclassification from the VVA model.
\begin{table}[ht]
\centering
\caption{Summary of simulation 4 results when $m=3$. }
\label{tab:s4m3}
\begin{tabular*}{0.9\textwidth}{@{\extracolsep{\fill}}lcccccc} 
\hline
Model & \thead{Missingness \\ mechanism} & \thead{Misclassification:\\ mean (s.d.) }& $\text{RMSE}(\hat{\matGamma})$ & $\text{RMSE}(\hat{\vecdelta})$ & $\text{RMSE}(\hat{\vecmu})$ & $\text{RMSE}(\hat{\matSigma})$\\
\hline
\multirow{2}{*}{}EEA & MAR &0.48 (0.10)  & 0.16 & 0.18& 0.55&0.19 \\ 
& \thead{State \& \\ Variable}& 0.29 (0.07) &0.09 & 0.11&0.38& 0.11\\ 

\hline
\multirow{2}{*}{}VVA& MAR & 0.52 (0.06)  & 0.20&0.20 & 0.59& 0.33\\ 
& \thead{State \& \\ Variable}& 0.39 (0.05) & 0.14&0.15 &0.46& 0.26\\ 

\hline
\multirow{2}{*}{}VEA    & MAR & 0.51 (0.08) &0.18 &0.17 & 0.57& 0.31\\ 
& \thead{State \& \\ Variable}& 0.32 (0.07) &0.10 & 0.12&0.41& 0.19\\ 
 \hline
\multirow{2}{*}{}EVA    & MAR & 0.50 (0.07) &0.19 & 0.20& 0.59& 0.25\\  
& \thead{State \& \\ Variable}& 0.35 (0.06) & 0.13& 0.12&0.44& 0.18\\ 
 \hline
\multirow{2}{*}{}VVI    & MAR & 0.42 (0.06) &0.14 &0.15 & 0.46& 0.22\\  
& \thead{State \& \\ Variable}& 0.30 (0.07) &0.10 & 0.11&0.39& 0.17\\ 
 \hline
\multirow{2}{*}{}VEI   & MAR & 0.41 (0.06) & 0.12&0.13 &0.44 & 0.22\\ 
& \thead{State \& \\ Variable}& 0.29 (0.06) & 0.09&0.10 &0.37&0.17 \\ 
 \hline
\multirow{2}{*}{}EVI    & MAR & 0.36 (0.06)& 0.10&0.11 &0.38 & 0.10\\ 
 & \thead{State \& \\ Variable}& 0.25 (0.05) & 0.08&0.09 &0.35&0.08 \\ 
 \hline
\multirow{2}{*}{}EEI   & MAR & 0.36 (0.06) & 0.11&0.11 &0.38 & 0.10\\ 
& \thead{State \& \\ Variable}& 0.25 (0.05) & 0.08&0.09 &0.35&0.08 \\ 
\hline
\end{tabular*}
\end{table}

\subsection{\textcolor{black}{Real Data}}
Results on two real datasets are presented below for all members of the CDGHMM family. When missing data and/or dropout is observed, the appropriate missingness mechanisms are tested and dropout is handled as detailed in Section \ref{EMcode}.
\subsubsection{Life Expectancy Data}

\begin{figure}[ht]
\hspace{1.3cm}
\centering
\captionsetup{labelfont={color=black},font={color=black}}
\includegraphics[width=0.7\linewidth]{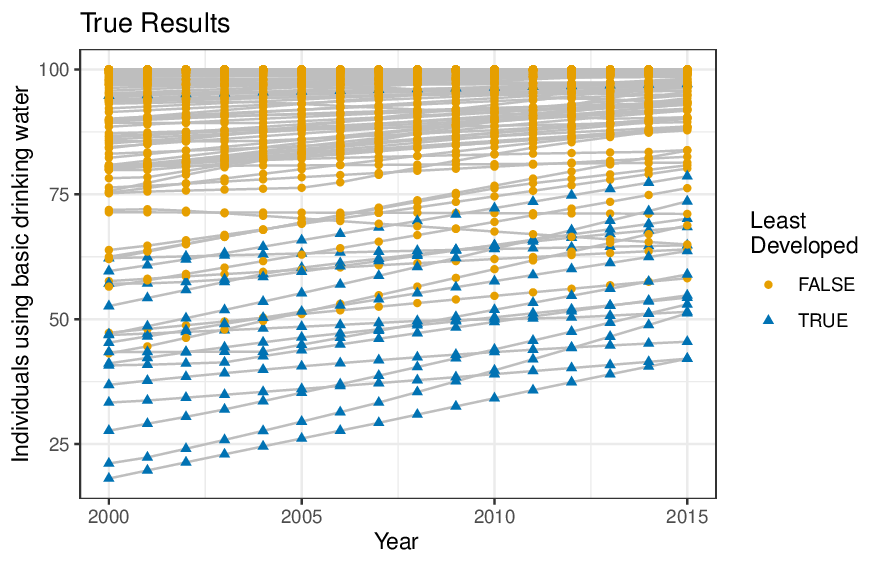}
\caption{True states of the life expectancy dataset.}
\label{fig:lifetrue}
\centering
\end{figure}

The Life Expectancy dataset \citep{lifedata} contains 13 continuous variables collected from 2000 to 2015 for 119 countries. It also includes a binary variable \texttt{Least Developed}, which equals TRUE if a country is considered least developed and FALSE otherwise (see Figure~\ref{fig:lifetrue} for a visualization). To test the accuracy of the algorithm, we assume a two-state HMM and compare the results to the variable \texttt{Least Developed}.

From Table \ref{tab:life.exp}, we see that both the BIC and the ICL select the VVA model, the worst performing model with a misclassification rate of 0.32. The VVI and VEI models are the best performing model in terms of the misclassification rate, at 0.18. \textcolor{black}{Because the VVA model corresponds to the traditional covariance update \eqref{eqn:thisone}, we see an improvement in state prediction from the proposed family over traditional HMMs.}
%\begin{singlespace}
\begin{table}[!ht]
\centering
\caption{CDGHMM results for the life expectancy dataset.}
\label{tab:life.exp}
\begin{tabular*}{0.5\textwidth}{@{\extracolsep{\fill}}cccc}
 \hline
 Model & \thead{Misclassification\\rate} & BIC & ICL \\ 
 \hline
 EEA & 0.19 & $-50752.8$ & $-46944.8$\\ 
 VVA & 0.32 & $-31956.3$ & $-28148.3$\\
 VEA & 0.20& $-47593.6$ & $-43785.6$\\
 EVA & 0.29 & $-36371.2$ & $-32563.2$\\
 VVI & 0.18 & $-50626.5$ & $-46818.5$\\
 VEI & 0.18 & $-52048.4$ & $-48240.4$\\
 EVI & 0.33 & $-55393.6$& $-51585.6$\\
 EEI & 0.26 & $-54688.8$ & $-50880.8$\\
 \hline
\end{tabular*}
\end{table}
%\end{singlespace}

\begin{figure}[!ht]% 
\captionsetup{labelfont={color=black},font={color=black}}
    \subfloat[\centering VVI model results]{{\includegraphics[width=0.48\linewidth]{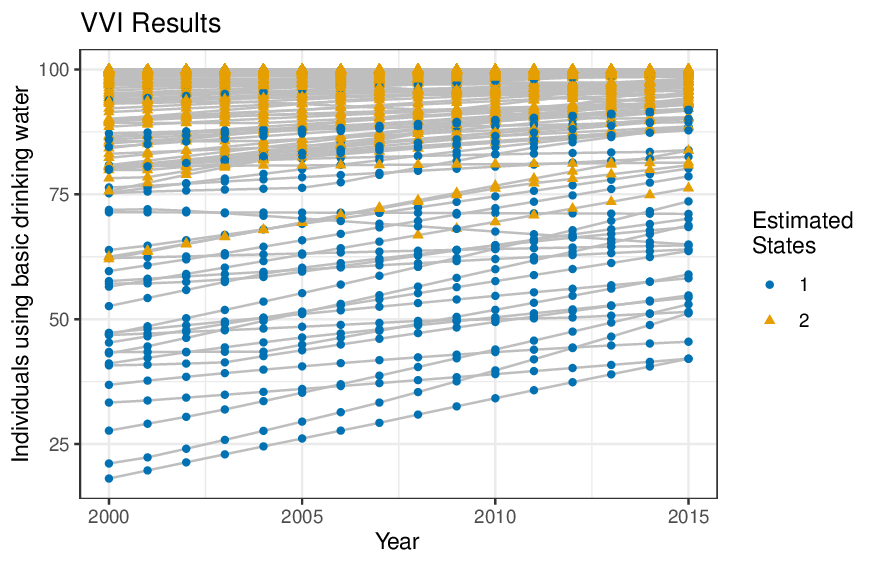}}}%
   \qquad
    \subfloat[\centering VVA model results]{{\includegraphics[width=0.48\linewidth]{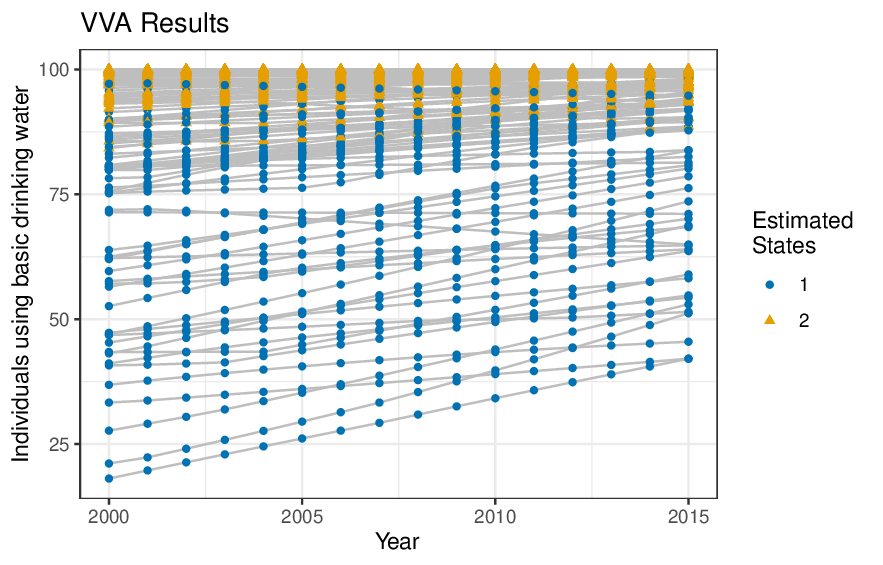}}}%
    \caption{Estimated states on the variable pertaining to individuals using basic drinking water from the life expectancy dataset.}
    \label{fig:vvivva}
    \end{figure}

\textcolor{black}{The estimated states are visualized for the VVA and VVI models on the variable pertaining to percentage of individuals using basic drinking water (Figure~\ref{fig:vvivva}). We see that although both models underestimate the number of countries in the more developed state, the VVA model underestimates said quantity to a greater degree.}

\subsubsection{Virtual Patient Model Dataset}
\begin{figure}[ht]
\hspace{1.3cm}
\centering
\captionsetup{labelfont={color=black},font={color=black}}
\includegraphics[width=0.6\linewidth]{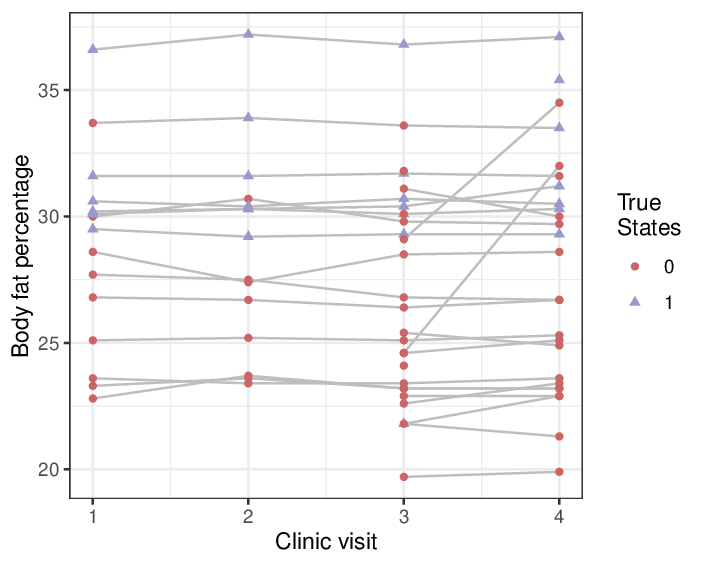}
\caption{True states of the virtual patient dataset.}
\label{fig:virtual}
\centering
\end{figure}

The Virtual Patient Model dataset \citep{konstantinos_deltouzos_2019_2670048} provides an overview of various physical, psychological, and cognitive health variables from 30 individuals between the ages of 70 and 85. Four time points are included in the study, including three individuals who dropout after the third time point. Missing data is present, primarily in the variable \texttt{Body fat percentage} and is seemingly related to state and time (see Figure \ref{fig:virtual}).

In the following analysis, nine physical health variables are used to fit a two-state HMM and the results are compared to an indicator variable detailing level of depression ($1$ if no to low levels of depression, $0$ otherwise). Table \ref{tab:virtual} contains the misclassification results for each member of the CDGHMM family and four missingness types: MAR, state-dependent, state- and variable-dependent, and state-, variable-, and time-dependent.
\begin{singlespace}
\begin{table}[ht]
\centering
\caption{Misclassification rates for CDGHMM family on the Virtual Patient Model dataset.}
\label{tab:virtual}
\begin{tabular*}{0.6\textwidth}{@{\extracolsep{\fill}}c c c c c c}
 \hline
 Model & MAR & State & \thead{State \& \\ Variable} & \thead{State, Variable, \\ \& Time} \\ 
 \hline
 EEA & 0.359 & 0.359 & 0.291 & 0.239 \\ 
 VVA &0.264 & 0.264 & 0.239 & 0.256 \\
  %\rowcolor{Gray}
 VEA & 0.274  & 0.274 & 0.274 & 0.274 \\
 EVA & 0.341  & 0.341 & 0.316 & 0.291 \\
 VVI &  0.299 & 0.299 & 0.274 & 0.256\\
 VEI &  0.299 & 0.299 & 0.274 & 0.256 \\
 EVI & 0.325  & 0.325 & 0.274 & 0.256 \\ 
 EEI &  0.333 & 0.333 & 0.282 & 0.282 \\ 
 \hline
\end{tabular*}
\end{table}
\end{singlespace}

From Table \ref{tab:virtual}, the EEA model with a state-, variable-, and time-dependent missingness mechanism and the VVA model with a state- and variable-dependent missingness mechanism perform the best in terms of misclassification rate. Interestingly, the missingness mechanisms appear to have different effects on performance for each model in the CDGHMM family, e.g., the EEA model and the VVA model. \textcolor{black}{The EEA model, under an MAR scheme, has a misclassification rate of 0.359. Under a state-, variable- and time-dependent missingness mechanism, this misclassification rate decreases to 0.239. For the VVA model, the decrease in misclassification rate when a non-random missingness mechanism is used is  smaller than for the EEA model.}

\section{Summary and Future Directions}
This research introduces a family of HMMs capable of accounting for parsimonious covariance structures in multivariate panel data. To do so, the proposed family uses a modified Cholesky-decomposed covariance structure with a series of optional constraints, resulting in eight models. \textcolor{black}{We see that this family aids in model estimation when the true covariance matrices are constrained but also when a high degree of state switching occurs} compared to a HMM with the traditional covariance update \eqref{eqn:thisone}. A modified EM algorithm is presented for model estimation for the CDGHMM family, with adaptations to handle non-random missingness and dropout. In doing so, better parameter estimates and predictive performance are achieved when compared to the MAR mechanism --- this is evidenced on both real and simulated data. 

For selection of the missingness mechanism, we urge users to perform an exploratory analysis to see if missingness appears to be related to time and/or any particular variables. This will then guide the selection of an appropriate missingness mechanism. From both the simulated and real data analyses, \textcolor{black}{we see that the proposed family and proposed missingness mechanisms work together to boost performance over traditional methods. As such, we suggest using the proposed methods together when missing data are present.} This work, however, highlights the need for effective selection criteria, which remains an open topic in the HMM literature. This work could easily be extended to HMMs with non-normal state-dependent processes and/or non-homogenous HMMs.

\section*{\textcolor{black}{Appendix}}
\appendix

\section {EEA Model}
\label{appEEA}
For the EEA model, $\matT_j = \matT$ and $\matD_j = \matD$. Through a similar process as in Sections~\ref{VVAmodel} and~\ref{EEImodel}, the expected value of the complete-data log-likelihood for the EEA model can be written
\begin{equation*}
    Q = C - \frac{N p}{2} \log 2\pi -  \frac{N}{2} \log |\matD| - \sum_{j = 1}^{m} \frac{n_j}{2} \text{tr}\{\mathbf{T} \mathbf{S}_j \mathbf{T}'\mathbf{D}^{-1}\},
\end{equation*}
where $C$ is a constant with respect to $\matT$ and $\matD$, $n_j = \sum_{i=1}^n\sum_{t=2}^{T} \hat{u}_{itj}$, and 
\[\matS_j = \frac{1}{n_j} \sum_{i=1}^{n} \sum_{t=1}^T \hat{u}_{itj} (\vecx_{it} - \vecmu_j) (\vecx_{it} - \vecmu_j)'. \]
Differentiating the expected complete-data log-likelihood with respect to $\matT$ and $\matD$, respectively, we get the following score functions:
\begin{align*}
    &S_1 (\matT, \matD) = \frac{\partial Q}{\partial \matT} =
    - \sum_{j=1}^m n_j (\matD^{-1})' \matT \matS_j,\\
    &S_2 (\matT, \matD) = \frac{\partial Q}{\partial \matD} = \frac{N}{2} \matD - \sum_{j=1}^m n_j \mathbf{T} \mathbf{S}_j \mathbf{T}'.
\end{align*}
Let $\phi_{ig}$ denote the lower triangular elements of $\matT$
and define $\Phi = \{ \phi_{ig}\}$ for $i > g$, with $i,g\in \{1,\ldots,p\}$. Consider solving $\text{LT}\{S_1 (\hat{\matT}, \matD)\}  \equiv [S_1 (\hat{\Phi}, \matD) ] = 0$ for $\hat{\Phi}$, where LT denotes the lower triangular part of a matrix and $\hat{\Phi}$ consists of $p-1$ systems of linear equations. In the case of the  $1 \times 1$ system, the solution is 
\begin{equation*}
    \sum_{j=1}^m n_j \left[ \frac{s_{11}^{(j)} \hat{\phi}_{21}}{d_{22}} + \frac{s_{21}^{(j)}}{d_{22}}  \right] = 0, 
\end{equation*}
and solving for $\hat{\phi}_{21}$ gives 
\begin{equation*}
     \hat{\phi}_{21} = - \frac{\sum_{j=1}^m \pi_j \left[s_{21}^{(j)} / d_{22} \right]}{\sum_{j=1}^m \pi_j \left[s_{11}^{(j)} / d_{22} \right]} = -\frac{\kappa_m^{21}}{\kappa_m^{11}},
\end{equation*}
where $\kappa_m^{ig} = \sum_{j=1}^m \pi_j \left[s_{ig}^{(j)} / d_{mm} \right]$ for convenience. Following this pattern, the general solution for an $r-1$ system of equations is 
\begin{equation}\label{eqn:app1}
    \begin{pmatrix}
    \hat{\phi}_{r1} \\
    \hat{\phi}_{r2} \\
    \vdots \\
    \hat{\phi}_{r,r-1}
    \end{pmatrix} 
    = - \begin{pmatrix}
    \kappa_r^{11} & \kappa_r^{21} & \hdots & \kappa_r^{r-1, 1} \\
    \kappa_r^{12} & \kappa_r^{22} & \hdots & \kappa_r^{r-1, 2} \\
    \vdots & \vdots & \ddots & \vdots \\
   \kappa_r^{1, r-1} & \kappa_r^{2, r-1} & \hdots & \kappa_r^{r-1, r-1} 
    \end{pmatrix}^{-1} 
     \begin{pmatrix}
    \kappa_r^{r1} \\
    \kappa_r^{r2} \\
     \vdots \\
     \kappa_r^{r,r-1}\\
     \end{pmatrix}
\end{equation}
for $r = 2,\ldots,p$. The $\hat{\phi}$ elements are then used in the $\hat{\matT}$ matrix as in Section \ref{VVAmodel}. Note that the $(r-1) \times (r-1)$ matrix in \eqref{eqn:app1} is symmetric because $\kappa_m^{ig} = \kappa_m^{gi}$. Solving diag$\left\{S_2(\hat{\matT}, \hat{\matD})\right\} = 0$ for $\hat{\matD}$ gives 
%\begin{equation*}
    $$\hat{\matD} = \sum_{j=1}^m \hat{\pi}_j \text{diag} \left\{ \hat{\matT} \matS_j \hat{\matT}' \right\}.$$
%\end{equation*}

\section{VEI Model}
\label{appVEI}
For the VEI model, $\matT_j = \matT_j$ and $\matD_j = \matD = d \mathbf{I}_p$. 
Again, through a similar process, the expected value of the complete-data log-likelihood for the VEI model can be written as 
\begin{equation*}
    Q = C + \sum_{j=1}^m n_j \log \pi_j + \frac{Np}{2} \log d^{-1} - \sum_{j=1}^m \frac{n_j d^{-1}}{2} \text{tr} \left\{ \matT \matS_j \matT' \right\},
\end{equation*}
where $C$ is a constant with respect to $\matT$ and $d$, $n_j = \sum_{i=1}^n\sum_{t=2}^{T} \hat{u}_{itj}$, and 
\[\matS_j = \frac{1}{n_j} \sum_{i=1}^{n} \sum_{t=1}^T \hat{u}_{itj} (\vecx_{it} - \vecmu_j) (\vecx_{it} - \vecmu_j)'. \]
When differentiating the expected complete-data log-likelihood with respect to $\matT$ and $d^{-1}$, respectively, we get the following score functions:
\begin{align*}
    &S_1 (\matT, d) = \frac{\partial Q}{\partial \matT} = - n_j d^{-1} (\matT_j \matS_j),\\
    &S_2 (\matT, d) = \frac{\partial Q}{\partial d^{-1}} = \frac{Npd}{2} - \sum_{j=1}^m \frac{n_j}{2} \text{tr} \left\{ \matT_j \matS_j \matT_j' \right\}.
\end{align*}
Again, let $\phi_{ig}^{(j)}$ denote the lower triangular elements of $\matT_j$
and define $\Phi_j = \{ \phi_{ig}^{(j)}\}$ for $i > g$, with $i,g\in \{1,\ldots,p\}$. Solving $S_1(\hat{\matT}_j, d) \equiv $ LT$\{S_1(\hat{\Phi}_j, d) \} = 0$ for $\hat{\Phi}_j$ presents us with $p-1$ systems of linear equations. In the case of the $1 \times 1 $ system, 
\begin{equation*}
    \frac{s_{11}^{(j)} \hat{\phi}_{21}^{(j)} }{d} + \frac{s_{21}^{(j)}}{d} = 0,
\end{equation*}
and so 
%\begin{equation*}
    $\hat{\phi}_{21}^{(j)} = -{s_{21}^{(j)}}/{s_{11}^{(j)}}$, 
%\end{equation*}
where $s_{ig}^{(j)}$ denotes the element $(i,g)$ of $\mathbf{S}_j$ and $d_i^{(j)}$ denotes the $i$th diagonal element of $\mathbf{D}_j$. For the general $(r-1) \times (r-1)$ case of system of equations,
\begin{equation*}
    \begin{pmatrix}
    \hat{\phi}_{r1}^{(j)} \\
    \hat{\phi}_{r2}^{(j)} \\
    \vdots \\
    \hat{\phi}_{r,r-1}^{(j)}
    \end{pmatrix} 
    = - \begin{pmatrix}
    s_{11}^{(j)} & s_{21}^{(j)} & \hdots & s_{r-1,1}^{(j)} \\
    s_{12}^{(j)} & s_{22}^{(j)} & \hdots & s_{r-1,2}^{(j)} \\
    \vdots & \vdots & \ddots & \vdots \\
    s_{1,r-1}^{(j)} & s_{2,r-2}^{(j)} & \hdots & s_{r-1,r-1}^{(j)}
    \end{pmatrix}^{-1} 
     \begin{pmatrix}
     s_{r1}^{(j)} \\
     s_{r2}^{(j)} \\
     \vdots \\
      s_{r,r-1}^{(j)} \\
     \end{pmatrix}
\end{equation*}
for $r = 2,\ldots,p$. Then, solving $S_2 (\hat{\mathbf{T}}_j, \hat{d})  = 0$ gives 
%\begin{equation*}
    $$\hat{d} = \frac{1}{p}\sum_{i=j}^m \hat{\pi}_j \text{tr}(\hat{\matT}_j \matS_j \hat{\matT}_j').$$
%\end{equation*}

\section {Selecting the Number of States}
\label{appCompSel}
\subsection{Simulation 2}
\begin{figure}[H] 
    \centering
    \subfloat[\centering BIC.\label{fig:panelA1}]{{\includegraphics[width=0.47\textwidth]{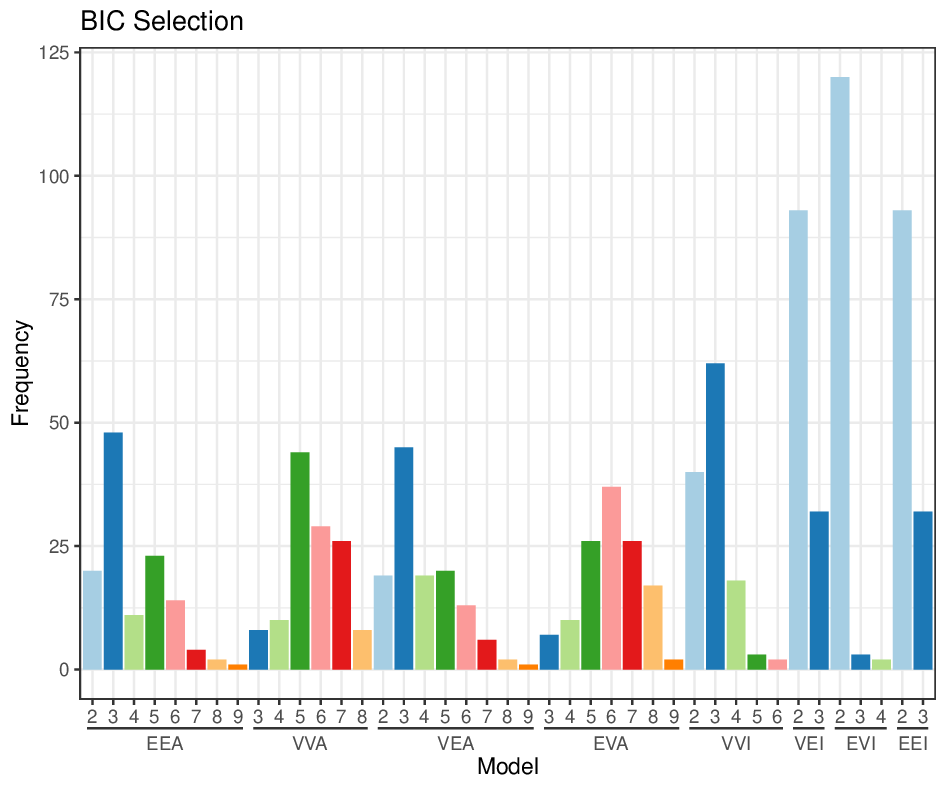}}}%
    \subfloat[\centering ICL. \label{fig:panelB1}]{{\includegraphics[width=0.47\textwidth]{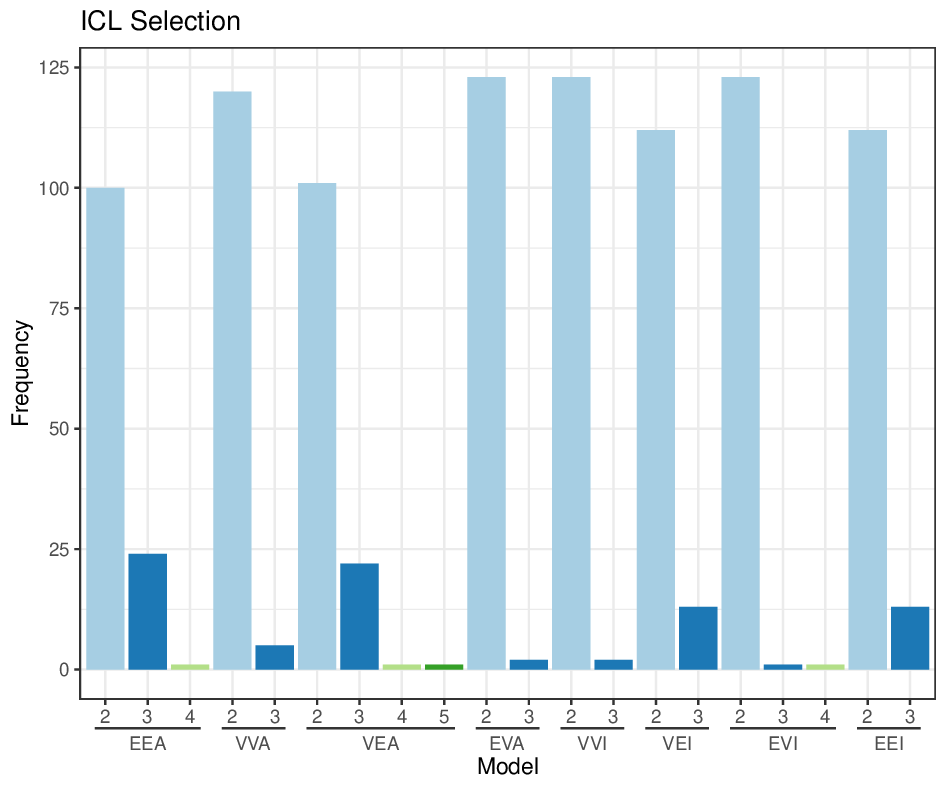}}}
    \newline
    \subfloat[\centering AIC. \label{fig:panelC1}]{{\includegraphics[width=0.47\textwidth]{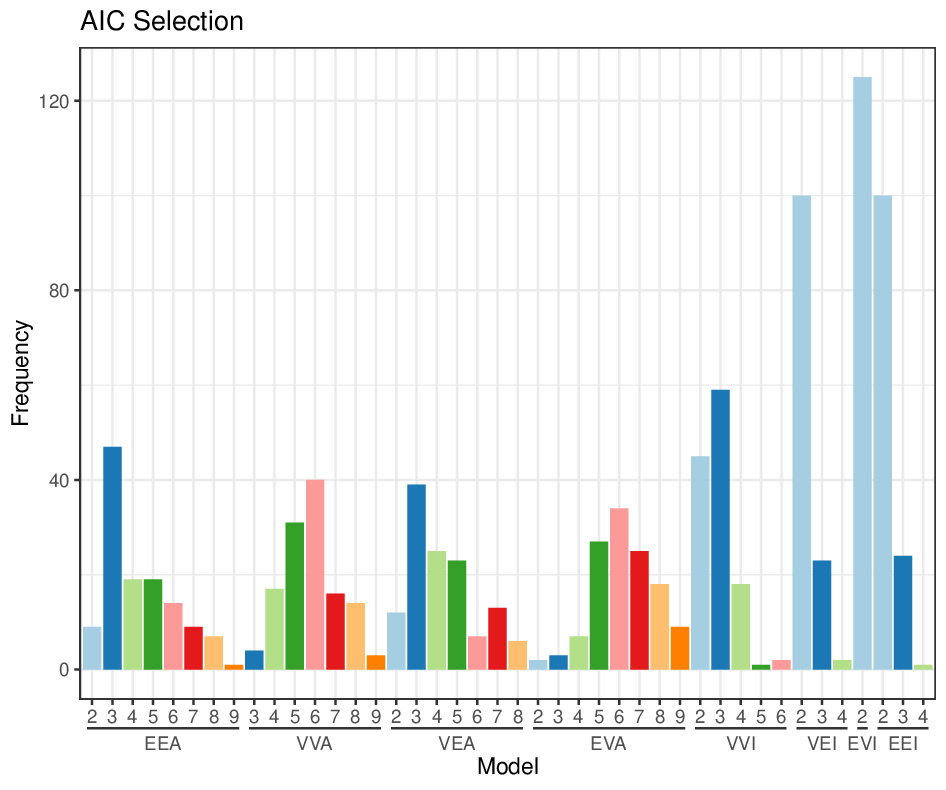}}} %
        \subfloat[\centering Average silhouette score. \label{fig:panelD1}]{{\includegraphics[width=0.47\textwidth]{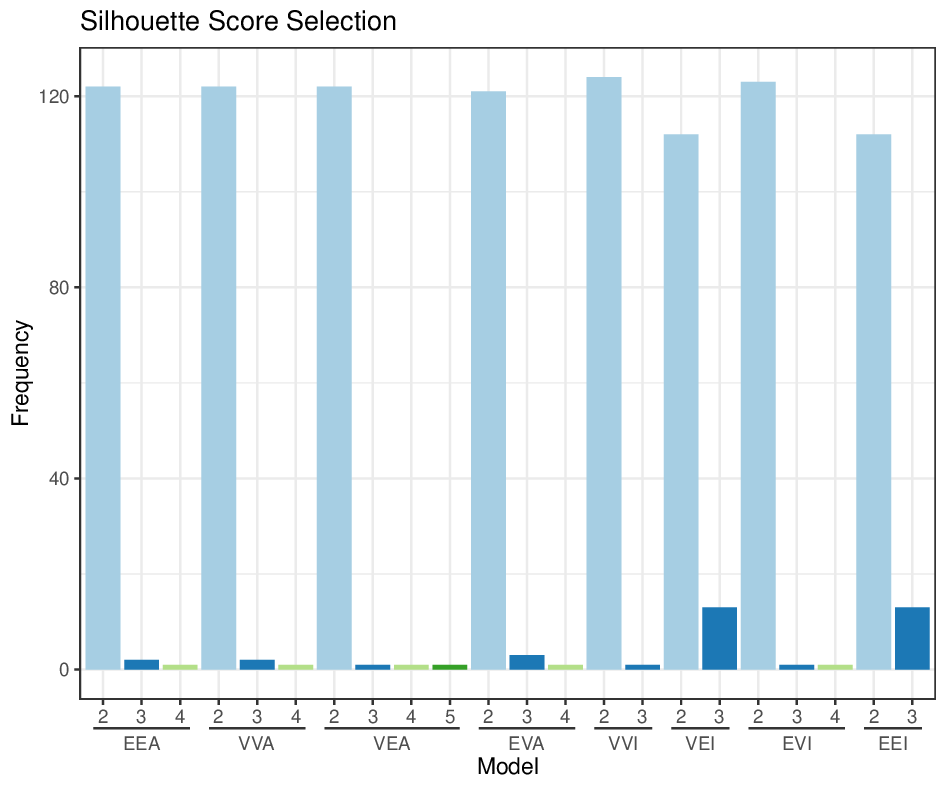}}}
    \caption{The frequency with which different numbers of states are selected using the BIC, ICL, AIC, and average silhouette score, respectively, based on Simulation~2.}
      \label{fig:panel7}
\end{figure}

Figure~\ref{fig:panel7} presents results from 125 iterations of Simulation~2 (see Section~\ref{sim2}), wherein we test BIC, ICL, AIC, and the average silhouette score on $m=2,\ldots,9$.

\subsection{Life Expectancy Dataset}
Table~\ref{tab:life_comp7} presents results from testing $m=2,\ldots,6$ on the life expectancy dataset. Note that results for $m>6$ are not presented due to issues around overfitting and singularities.
\begin{singlespace}
\begin{table}[ht]
\centering
\caption{Selected number of states for each selection criterion and model.}
\label{tab:life_comp7}
\begin{tabular*}{0.6\textwidth}{@{\extracolsep{\fill}}c c c c c}
 \hline
 Model & BIC & ICL & AIC & \thead{Average \\ Silhouette}\\ 
 \hline
 EEA & 6 & 6 & 6 & 2\\ 
 VVA &6 & 6& 6 & 2 \\
 VEA & 6  & 6 & 6 & 2 \\
 EVA & 6  & 6 & 6 & 2 \\
 VVI & 6 & 6 & 6 & 2\\ 
 VEI &6 & 6 & 6& 2 \\
 EVI & 6  & 6 & 6 & 3 \\
EEI & 6  & 6 & 6 & 3 \\
 \hline
\end{tabular*}
\end{table}
\end{singlespace}

\section*{Acknowledgements}
 This work was supported by a Discovery Grant from the Natural Sciences and Engineering Research Council of Canada, the Canada Research Chairs program, and a Dorothy Killam Fellowship.


\begin{thebibliography}{}

\bibitem[\protect\citeauthoryear{Akaike}{Akaike}{2011}]{akaike2011}
Akaike, H. (2011).
\newblock Akaike’s information criterion.
\newblock {\em International Encyclopedia of Statistical Science.\/} 25--25.

\bibitem[\protect\citeauthoryear{Anderson}{Anderson}{1984}]{anderson2003introduction}
Anderson, T.~W. (1984).
\newblock {\em An Introduction to Multivariate Statistical Analysis\/} (2 ed.).
\newblock New Jersey: Wiley.

\bibitem[\protect\citeauthoryear{Baum}{Baum}{1972}]{baum1972inequality}
Baum, L.~E. (1972).
\newblock An inequality and associated maximization technique in statistical
  estimation for probabilistic functions of {M}arkov processes.
\newblock {\em Inequalities\/}~{\em 3\/}(1), 1--8.

\bibitem[\protect\citeauthoryear{Baum, Petrie, Soules, and Weiss}{Baum
  et~al.}{1970}]{baum1970maximization}
Baum, L.~E., T.~Petrie, G.~Soules, and N.~Weiss (1970).
\newblock A maximization technique occurring in the statistical analysis of
  probabilistic functions of {M}arkov chains.
\newblock {\em The Annals of Mathematical Statistics\/}~{\em 41\/}(1),
  164--171.

\bibitem[\protect\citeauthoryear{Beno\^{i}t}{Beno\^{i}t}{1924}]{benoit1924note}
Beno\^{i}t, C. (1924).
\newblock Note sur une m{\'e}thode de r{\'e}solution des {\'e}quations normales
  provenant de l'application de la m{\'e}thode des moindres carr{\'e}s {\`a} un
  syst{\`e}me d'{\'e}quations lin{\'e}aires en nombre inf{\'e}rieur {\`a} celui
  des inconnues ({P}roc{\'e}d{\'e} du {C}ommandant {C}holesky).
\newblock {\em Bulletin G{\'e}od{\'e}sique\/}~{\em 2\/}(1), 67--77.

\bibitem[\protect\citeauthoryear{Biernacki, Celeux, and Govaert}{Biernacki
  et~al.}{2000}]{biernacki2000assessing}
Biernacki, C., G.~Celeux, and G.~Govaert (2000).
\newblock Assessing a mixture model for clustering with the integrated
  completed likelihood.
\newblock {\em IEEE Transactions on Pattern Analysis and Machine
  Intelligence\/}~{\em 22\/}(7), 719--725.

\bibitem[\protect\citeauthoryear{Celeux and Durand}{Celeux and
  Durand}{2008}]{celeux2008selecting}
Celeux, G. and J.-B. Durand (2008).
\newblock Selecting hidden {M}arkov model state number with cross-validated
  likelihood.
\newblock {\em Computational Statistics\/}~{\em 23}, 541--564.

\bibitem[\protect\citeauthoryear{Deltouzos}{Deltouzos}{2019}]{konstantinos_deltouzos_2019_2670048}
Deltouzos, K. (2019, May).
\newblock Aggregated virtual patient model dataset.

\bibitem[\protect\citeauthoryear{Dempster, Laird, and Rubin}{Dempster
  et~al.}{1977}]{dempster1977maximum}
Dempster, A.~P., N.~M. Laird, and D.~B. Rubin (1977).
\newblock Maximum likelihood from incomplete data via the {EM} algorithm.
\newblock {\em Journal of the Royal Statistical Society: Series~B\/}~{\em
  39\/}(1), 1--22.

\bibitem[\protect\citeauthoryear{du~Roy~de Chaumaray and Marbac}{du~Roy~de
  Chaumaray and Marbac}{2023}]{du2023clustering}
du~Roy~de Chaumaray, M. and M.~Marbac (2023).
\newblock Clustering data with non-ignorable missingness using semi-parametric
  mixture models assuming independence within components.
\newblock {\em Advances in Data Analysis and Classification\/}~{\em 17},
  1081--1122.

\bibitem[\protect\citeauthoryear{Eirola, Lendasse, Vandewalle, and
  Biernacki}{Eirola et~al.}{2014}]{eirola2014mixture}
Eirola, E., A.~Lendasse, V.~Vandewalle, and C.~Biernacki (2014).
\newblock Mixture of {G}aussians for distance estimation with missing data.
\newblock {\em Neurocomputing\/}~{\em 131}, 32--42.

\bibitem[\protect\citeauthoryear{Fr{\"u}hwirth-Schnatter}{Fr{\"u}hwirth-Schnatter}{2006}]{fruhwirth2006finite}
Fr{\"u}hwirth-Schnatter, S. (2006).
\newblock {\em Finite Mixture and {M}arkov Switching Models}.
\newblock New York: Springer.

\bibitem[\protect\citeauthoryear{Ghahramani and Jordan}{Ghahramani and
  Jordan}{1994}]{ghahramani1995learning}
Ghahramani, Z. and M.~I. Jordan (1994).
\newblock Learning from incomplete data.
\newblock Technical Report AIM-1509, Massachusetts Institute of Technology,
Cambridge, MA.

\bibitem[\protect\citeauthoryear{Hasan and Sneddon}{Hasan and
  Sneddon}{2009}]{hasan2009zero}
Hasan, M.~T. and G.~Sneddon (2009).
\newblock Zero-inflated {P}oisson regression for longitudinal data.
\newblock {\em Communications in Statistics--Simulation and Computation\/}~{\em
  38\/}(3), 638--653.
  
 \bibitem[\protect\citeauthoryear{Hung, Wang, Zarnitsyna, Zhu, and Wu}{Hung et~al.}{2013}]{hung2013}
Hung, Y., Y.~Wang, V.~ Zarnitsyna, C.~ Zhu and C.F.J.~Wu (2013).
\newblock Hidden Markov models with applications in cell adhesion experiments.
\newblock {\em Journal of the American Statistical Association\/}~{\em 108\/}(504),
  1469--1479.

\bibitem[\protect\citeauthoryear{Hunt and Jorgensen}{Hunt and
  Jorgensen}{2003}]{hunt2003mixture}
Hunt, L. and M.~Jorgensen (2003).
\newblock Mixture model clustering for mixed data with missing information.
\newblock {\em Computational Statistics and Data Analysis\/}~{\em 41\/}(3-4),
  429--440.

\bibitem[\protect\citeauthoryear{Kuha, Katsikatsou, and Moustaki}{Kuha
  et~al.}{2018}]{kuha2018latent}
Kuha, J., M.~Katsikatsou, and I.~Moustaki (2018).
\newblock Latent variable modelling with non-ignorable item non-response:
  multigroup response propensity models for cross-national analysis.
\newblock {\em Journal of the Royal Statistical Society: Series~A\/}~{\em
  181\/}(4), 1169--1192.
  
 \bibitem[\protect\citeauthoryear{Lin and Song}{Lin and Song}{2022}]{lin22}
Lin, Y. and X.~Song (2022).
\newblock Order selection for regression-based hidden Markov model.
\newblock {\em Journal of Multivariate Analysis\/}~{\em
  192\/}, 105061.

\bibitem[\protect\citeauthoryear{Maruotti}{Maruotti}{2011}]{maruotti2011mixed}
Maruotti, A. (2011).
\newblock Mixed hidden {M}arkov models for longitudinal data: An overview.
\newblock {\em International Statistical Review\/}~{\em 79\/}(3), 427--454.

\bibitem[\protect\citeauthoryear{McNicholas}{McNicholas}{2007}]{mcnicholas07}
McNicholas, P.~D. (2007).
\newblock {\em Topics in Unsupervised Learning}.
\newblock Ph.\ D. thesis, Trinity College Dublin.

\bibitem[\protect\citeauthoryear{McNicholas and Murphy}{McNicholas and
  Murphy}{2010}]{mcnicholas2010model}
McNicholas, P.~D. and T.~B. Murphy (2010).
\newblock Model-based clustering of longitudinal data.
\newblock {\em Canadian Journal of Statistics\/}~{\em 38\/}(1), 153--168.

\bibitem[\protect\citeauthoryear{Neal, Sochaniwsky, and McNicholas}{Neal et~al.}
{2024}]{neal24a}
Neal, M.~R., A.~A. Sochaniwsky, and P.~D. McNicholas (2024).
\newblock CDGHMM: Hidden Markov models for multivariate panel data.
\newblock R package version 0.1.0.


\bibitem[\protect\citeauthoryear{Pandolfi, Bartolucci, and Pennoni}{Pandolfi
  et~al.}{2023}]{pandolfi2023hidden}
Pandolfi, S., F.~Bartolucci, and F.~Pennoni (2023).
\newblock A hidden {M}arkov model for continuous longitudinal data with missing
  responses and dropout.
\newblock {\em Biometrical Journal\/}~{\em 65\/}(5), 2200016.

\bibitem[\protect\citeauthoryear{Pohle, Langrock, Van~Beest, and Schmidt}{Pohle
  et~al.}{2017}]{pohle2017selecting}
Pohle, J., R.~Langrock, F.~M. Van~Beest, and N.~M. Schmidt (2017).
\newblock Selecting the number of states in hidden {M}arkov models: pragmatic
  solutions illustrated using animal movement.
\newblock {\em Journal of Agricultural, Biological and Environmental
  Statistics\/}~{\em 22}, 270--293.

\bibitem[\protect\citeauthoryear{Popov, Gultyaeva, and Uvarov}{Popov
  et~al.}{2016}]{popov2016training}
Popov, A.~A., T.~A. Gultyaeva, and V.~E. Uvarov (2016).
\newblock Training hidden {M}arkov models on incomplete sequences.
\newblock In {\em 2016 13th International Scientific-Technical Conference on
  Actual Problems of Electronics Instrument Engineering (APEIE)}, Volume~2,
  pp.\  317--320. IEEE.

\bibitem[\protect\citeauthoryear{Pourahmadi}{Pourahmadi}{1999}]{pourahmadi1999joint}
Pourahmadi, M. (1999).
\newblock Joint mean-covariance models with applications to longitudinal data:
  Unconstrained parameterisation.
\newblock {\em Biometrika\/}~{\em 86\/}(3), 677--690.

\bibitem[\protect\citeauthoryear{Pourahmadi}{Pourahmadi}{2000}]{pourahmadi2000maximum}
Pourahmadi, M. (2000).
\newblock Maximum likelihood estimation of generalised linear models for
  multivariate normal covariance matrix.
\newblock {\em Biometrika\/}~{\em 87\/}(2), 425--435.

\bibitem[{{R Core Team}(2023)}]{R23}
R Core Team (2023).
\newblock R: A Language and Environment for Statistical Computing.
\newblock R Foundation for Statistical Computing, Vienna, Austria.
\url{https://www.R-project.org/}

\bibitem[\protect\citeauthoryear{Rousseeuw}{Rousseeuw}{1987}]{Rousseeuw1987}
Rousseeuw, P.~J. (1987).
\newblock Silhouettes: a graphical aid to the interpretation and validation of cluster analysis.
\newblock {\em Journal of Computational and Applied Mathematics\/}~{\em 20\/}, 53--65.

\bibitem[\protect\citeauthoryear{Schwarz}{Schwarz}{1978}]{schwarz1978estimating}
Schwarz, G. (1978).
\newblock Estimating the dimension of a model.
\newblock {\em The Annals of Statistics\/}, 461--464.

\bibitem[\protect\citeauthoryear{Speekenbrink and Visser}{Speekenbrink and
  Visser}{2021}]{speekenbrink2021ignorable}
Speekenbrink, M. and I.~Visser (2021).
\newblock Ignorable and non-ignorable missing data in hidden {M}arkov models.
\newblock arXiv preprint arXiv:2109.02770.

\bibitem[\protect\citeauthoryear{Sportisse, Marbac, Biernacki, Boyer, Celeux,
  Josse, and Laporte}{Sportisse et~al.}{2021}]{sportisse2021model}
Sportisse, A., M.~Marbac, C.~Biernacki, C.~Boyer, G.~Celeux, J.~Josse, and
  F.~Laporte (2021).
\newblock Model-based clustering with missing not at random data.
\newblock arXiv preprint arXiv:2112.10425.

\bibitem[\protect\citeauthoryear{Sutradhar}{Sutradhar}{2003}]{sutradhar2003overview}
Sutradhar, B.~C. (2003).
\newblock An overview on regression models for discrete longitudinal responses.
\newblock {\em Statistical Science\/}~{\em 18\/}(3), 377--393.

\bibitem[\protect\citeauthoryear{vrec99}{vrec99}{2022}]{lifedata}
vrec99 (2022).
\newblock Life expectancy 2000-2015.
\newblock
  \url{https://www.kaggle.com/datasets/vrec99/life-expectancy-2000-2015}.

\bibitem[\protect\citeauthoryear{Welch}{Welch}{2003}]{welch2003hidden}
Welch, L.~R. (2003).
\newblock Hidden {M}arkov models and the {B}aum-{W}elch algorithm.
\newblock {\em IEEE Information Theory Society Newsletter\/}~{\em 53\/}(4),
  10--13.
  
\bibitem[\protect\citeauthoryear{Zou, Lin, Song}{Zou
  et~al.}{2024}]{zou24}
Zou, Y., Y.~Lin and X.~Song (2024).
\newblock Bayesian heterogeneous hidden Markov models with an unknown number of states.
\newblock {\em Journal of Computational and Graphical Statistics\/}~{\em 33\/}(1), 15--24.

\bibitem[\protect\citeauthoryear{Zucchini and MacDonald}{Zucchini and
  MacDonald}{2009}]{zucchini2009hidden}
Zucchini, W. and I.~L. MacDonald (2009).
\newblock {\em Hidden {M}arkov Models for Time Series: An Introduction Using
  R}.
\newblock Boca Raton: Chapman and Hall/CRC Press.

\end{thebibliography}
\end{document}